\documentclass[twocolumn]{aastex63}

\usepackage[T1]{fontenc}
\usepackage{bm}        
\usepackage{amssymb, amsmath}   
\usepackage{cancel}
\usepackage{color}
\usepackage{verbatim}
\usepackage{wrapfig,xkeyval,svg}
\usepackage{tikz,hyperref}
\usepackage{float}
\usepackage{placeins} 
\usepackage{soul}
\UseRawInputEncoding
\DeclareMathAlphabet{\mathsfit}{\encodingdefault}{\sfdefault}{m}{sl}
\SetMathAlphabet{\mathsfit}{bold}{\encodingdefault}{\sfdefault}{bx}{sl}

\newcommand{\eqbreak}[1][2]{\\&\hskip#1em}

\makeatletter
\newlength{\sfp@hseplen}\newlength{\sfp@vseplen}
\define@cmdkey{subfigpos}[sfp@]{pos}[ul]{}
\define@cmdkey{subfigpos}[sfp@]{font}[\small]{}
\define@cmdkey{subfigpos}[sfp@]{vsep}[2\baselineskip]{\setlength{\sfp@vseplen}{\sfp@vsep}}
\define@cmdkey{subfigpos}[sfp@]{hsep}[10pt]{\setlength{\sfp@hseplen}{\sfp@hsep}}
\newcommand{\subfigimg}[3][,]{%
  \setkeys{Gin,subfigpos}{pos,font,vsep,hsep,#1}
  \setbox1=\hbox{\includegraphics{#3}}
  \ifnum\pdfstrcmp{\sfp@pos}{ul}=0
    \leavevmode\rlap{\usebox1}
    \rlap{\hspace*{\sfp@hsep}\raisebox{\dimexpr\ht1-\sfp@vsep}{\sfp@font{#2}}}
    \phantom{\usebox1}
  \else\ifnum\pdfstrcmp{\sfp@pos}{ur}=0
    \leavevmode\usebox1
    \llap{\raisebox{\dimexpr\ht1-\sfp@vsep}{\sfp@font{#2}}\hspace*{\sfp@hsep}}
  \else\ifnum\pdfstrcmp{\sfp@pos}{lr}=0
    \leavevmode\usebox1
    \llap{\raisebox{\sfp@vsep}{\sfp@font{#2}}\hspace*{\sfp@hsep}}
  \else
    \leavevmode\rlap{\usebox1}
    \rlap{\hspace*{\sfp@hseplen}\raisebox{\sfp@vsep}{\sfp@font{#2}}}
    \phantom{\usebox1}
  \fi\fi\fi
}

\shorttitle{Particle Acceleration and Radiative Output in Relativistic SBLs}
\shortauthors{Chand \& B\"ottcher}

\begin{document}
\defcitealias{Chand2021}{Paper I}

\title{Inverse Compton Emission and Cooling of Relativistic Particles Accelerated at Shear Boundary Layers in Relativistic Jets}

\correspondingauthor{Tej Chand}
\email{chandtej11@gmail.com}

\author[0000-0002-1833-3749]{Tej Chand}
\affiliation{Centre for Space Research \\
North-West University \\
Potchefstroom, 2520, South Africa}

\author[0000-0002-8434-5692]{Markus B\"ottcher}
\affiliation{Centre for Space Research \\
North-West University \\
Potchefstroom, 2520, South Africa}

\begin{abstract}
Both observational evidence and theoretical considerations from MHD simulations of jets suggest that the relativistic jets of active galactic nuclei (AGN) are radially stratified, with a fast inner spine surrounded by a slower-moving outer sheath. The resulting relativistic shear layers are a prime candidate for the site of relativistic particle acceleration in the jets of AGN and gamma-ray bursts (GRBs). In this article, we present outcomes of particle-in-cell simulations of magnetic-field generation and particle acceleration in the relativistic shear boundary layers (SBLs) of jets in AGN and GRBs. We investigate the effects of inverse Compton cooling on relativistic particles that are accelerated in the SBLs of relativistic jets including the self-consistent calculation of the radiation spectrum produced by inverse Compton scattering of relativistic electrons in an isotropic external soft photon field. We find that the Compton cooling can be substantial, depending on the characteristic energy (blackbody temperature and energy density) of the external radiation field. The produced Compton emission is highly anisotropic and more strongly beamed along the jet direction than the characteristic $1/\Gamma$ pattern expected from intrinsically isotropic emission in the comoving frame of an emission region moving along the jet with bulk Lorentz factor $\Gamma$. We suggest that this may resolve the long-standing problem of the Doppler Factor Crisis.
\end{abstract}

\keywords{relativistic jets: shear boundary layers, shear acceleration, radiation drag, radiation mechanism}


\section{Introduction} \label{sec:intro}

The existence of jets associated with accretion onto compact objects is ubiquitous in astrophysical systems. Those collimated relativistic outflows emanating from a variety of astrophysical sources like a stellar mass compact object, such as a white dwarf (WD), a neutron star (NS), or a black hole (BH), in close binaries, as well as jets from gamma-ray bursts (GRBs) can travel over pc scales whereas those from some active galactic nuclei (AGNs) extend over kpc scales without showing significant loss of momentum and kinetic energy. Many processes in and properties of relativistic jets, like jet composition, collimation, particle acceleration, stabilization, mass-loading mechanism, and radiative mechanisms, are still poorly understood. A recent review on relativistic jets can be found in, e.g., \citet{2017_review_Markus}.

Both observational evidence and theoretical considerations from Magneto-Hydrodynamic (MHD) simulations indicate that astrophysical jets have radial or transverse structure \citep[e.g.,][]{2007..mhd,2006..mhd,10.1093/mnras/237.2.411,2004ApJ...600..127G}. When the powerful jet from the central engine travels through the interstellar plasma, sharp boundary layers can be formed due to the velocity difference between the jet and the surrounding plasma. Also, the jet may accelerate to different intrinsic bulk speeds at different radial distances from the jet axis. The ensuing shear boundary layers (SBLs) are promising sites for the particle acceleration \citep[e.g.,][]{stawarz_ostrowski_2002,10.1111/j.1745-3933.2009.00707.x,2013(a),Chand:2019plf}. The formation of SBLs is expected in a variety of astrophysical environments like AGN jets \citep[e.g.,][]{2012_Alves,2013(a),galaxies7030078} and also in the jets involving ultra-relativistic outflows of GRBs \citep[e.g.,][]{2000}. The majority of such astrophysical systems are collisionless, wherein the electron mean free path is much larger than the system size, requiring Particle-in-Cell (PiC) methods to study the kinetic effects. The angular variations in the jet Lorentz factor indicated from the recent observations of the afterglow of the neutron star-neutron star merger GW170817 \citep{2017_GW} also support the picture of a structured jet in which the Lorentz factor of the jet and its energetics vary with the angle from the central axis \citep[e.g.,][]{ 2017_Lamb,10.1093/mnras/staa3026}.  \citet{2005A&A...432..401G} proposed a spine-layer structure of blazar and radio galaxy jets in order to interpret the longstanding issue of the Bulk Lorentz Factor Crisis (BLFC) in some blazar jets \citep[e.g.,][]{2006,2010}. Recent studies of SBLs using Particle-in-Cell \citep{106800} simulations of relativistic jets have demonstrated that from initially unmagnetized plasma electric and magnetic fields can be efficiently generated leading to particle acceleration \citep[e.g.,][]{2012_Alves, 2013(b), galaxies7030078}. Recent studies \citep[e.g.,][]{2013_Grismayer,2013_Grismayer_second,2014_Alves} have provided theoretical predictions and a physical picture of plasma instabilities like Kelvin-Helmholtz instability, KHI \citep{1961hhs..book.....C}, or the Weibel instability \citep{PhysRevLett.2.83}, that can develop within shear layers. These studies suggest that the development of magnetic fields in SBLs is due to those plasma instabilities. Although the shear interface, in the hydrodynamic limit, is unstable in the case of the classical KHI, the instability is suppressed by the longitudinal B-field in the presence of ambient magnetic fields \citep{1961hhs..book.....C}. Thus, the shear interface in the kpc jets of AGN is stable against the KHI. The impact of the transverse dynamics of the KHI in unmagnetized shearing flow, which is not accounted for in 2D simulation, has recently been explored through the use of 3D PiC simulations \citep[e.g.,][]{2012_Alves}. While the longitudinal dynamics of the KHI hold significance in both subrelativistic and relativistic regimes, the transverse dynamics become particularly notable in the relativistic regime, in contrast to the subrelativistic regime where longitudinal dynamics dominate. In relativistic regimes, the complete 3D evolution of the KHI is governed by transverse dynamics, characterized by the aggregation of a Weibel-like electron bunching process. This results in the creation of electron current filaments, ultimately driving the acceleration of electrons across the shear interface.

The most widely accepted acceleration mechanism in the scenario of relativistic jets is the first-order Fermi acceleration mechanism. Considerable attention is given to shock acceleration as a particle acceleration mechanism in relativistic jets. However, the Fermi process is also possible without a shock where scattering centres flow at different speeds despite the fact that the flows are parallel. The particles attain different velocities between the fast-moving spine and the slow-moving layer. This mechanism, called shear acceleration in shear flows, occurs as a result of energetic particles encountering different local velocities in the collisionless background flow \citep[e.g.,][]{2004_shear}. Recent high-resolution studies of extragalactic jets indicate that the first-order Fermi mechanism alone cannot fully reckon the detection of extended high energy emission \citep[e.g.,][]{rieger_duffy_2006}. The strong synchrotron cooling process which is expected in the shock acceleration scenarios is not observed, e.g., in the case of the quasar 3C 273 \citep[e.g.,][]{2001_Jester}, suggesting that there is the need for a continuous re-acceleration mechanism in the relativistic jets to resolve these issues. Thus, shear layer acceleration is likely to be significant in high-energy astrophysical phenomena. The energetic particles may sample the velocity difference while moving across the shear flow. Magnetic field inhomogeneities embedded in different layers of the shear flow scatter energetic particles, leading to energy gain due to bulk velocity differences across the shear flow \citep[e.g.,][]{2017_Ruo}. This eventually converts the bulk motion kinetic energy of the background flows to nonthermal particle energies. Previous studies have indicated that the efficiency of the shear acceleration depends on the strength of the velocity shear and hence the shear acceleration is employed more efficiently in relativistic jets than in non-relativistic ones \citep[e.g.,][]{1989ApJ...340.1112W,1990A&A...238..435O}.

The spectral energy distributions (SEDs) of blazars consist typically of two broad, non-thermal components. The lower energy bump, attributed to the synchrotron emission from relativistic leptons, is observed in the radio through optical/UV bands and in some cases up to X-rays. In leptonic models, the higher energy bump is attributed to the Inverse Compton (IC) scattering by the same leptons of the synchrotron photons (SSC) or External photons (EC). High energy particles energized at SBLs across magnetic field lines have an anisotropic momentum distribution with efficient emission of Synchrotron radiation \citep[e.g.,][]{2013(a)}. Several investigations employing PiC simulations have examined the influence of synchrotron cooling \citep[e.g.,][]{Hakobyan_2019} and IC cooling \citep[e.g.,][]{10.1093/mnrasl/sly157} on relativistic jet particles within the framework of magnetic reconnection, including the resulting radiation spectra. However, despite the potential for radiative cooling to significantly influence particle dynamics, the effect of radiation drag on radiating particles and the resulting radiation spectra largely remain unexplored in studies of particle acceleration at SBLs in jets.

In this study, we investigate the acceleration mechanism and radiation properties of relativistic jets' SBLs. We demonstrate that SBL turbulence generated by self-generated electromagnetic fields produces high-energy particles. Furthermore, we calculate the radiation spectra by considering the IC scattering of relativistic electrons in an angle-averaged and angle-dependent soft photon field, including the impact of IC cooling on particle dynamics and energetics. In section \ref{sec:setup}, the physical and simulation setup are discussed with the initial simulation parameters. In section \ref{sec: ICS-theory}, we discuss the theoretical background of an IC scattering of an external blackbody photon field and how we evaluate the radiation spectra. In section \ref{sec:results}, we present our PiC simulation results of self-generated electric and magnetic fields, particle spectra, and radiative output. In section \ref{sec:conclusion}, we summarize the main results.
\section{Model Setup \label{sec:setup}}
In this section, we present the physical and simulation setups used for our study of particle acceleration and inverse Compton emission in spine-sheath plasma jet structures of relativistic jets.
\bigskip
\subsection{Physical Model \label{subsec:physical_setup}}
In our simulations, we examine the mechanisms behind particle acceleration and radiation in SBLs of relativistic jets using a spine-sheath plasma jet structure. The jet consists of two parts moving in opposite directions along the $\rm x$-axis with the same velocity and Lorentz factor of $\Gamma = 15$. The spine moves in the positive $\rm x$-direction while the sheath moves in the negative $\rm x$-direction. The plasma in this simulation is initially unmagnetized ($\rm B=0$) and has a temperature of $\mathrm{2.5 ~ keV}$. Table \ref{table:tab_ph_parameter} contains a list of the physical parameters that we use in our simulations.

\begin{deluxetable}{lccr}\label{table:tab_ph_parameter} 
\tablecaption{Table of physical parameters} 
\tablehead{ 
    \colhead{Physical parameters} &
    \colhead{Values} 
}
\startdata 
    Initial magnetic field (B) & 0   \\
   Ion temperature ($\mathrm{K_ BT_i}$) & 2.5 keV \\
   Electron temperature ($\mathrm{K_BT_e}$) & 2.5 keV \\
   Bulk Lorentz factor of the spine ($\Gamma_{\rm sp}$) & 15 \\
   Bulk Lorentz factor of the sheath ($\Gamma_{\rm sh}$) & 15 \\
\enddata
\end{deluxetable}

\bigskip
\subsection{Simulation Setup \label{subsec:sim_setup}}
To study the particle acceleration and radiation mechanism in SBLs of relativistic jets, we perform fully kinetic 2.5D electromagnetic Particle-in-Cell (PiC) simulations using the \texttt{TRISTAN-MP} code \citep{2005}. The rectangular computational domain we consider in the $\rm XY$ plane has periodic boundaries with a total of $\mathrm{L_x=1024}$ grid points in the $\rm x$-direction and $\mathrm{L_y=2L_x}$ in the $\rm y$-direction. All distances are measured in units of electron skin depth, $\mathrm{d_e = c/\omega_{p,e}}$, where $\rm c$ is the speed of light and $\mathrm{\omega_{p,e} = \sqrt{4 \pi n e^2/m_e}}$ is the electron plasma frequency ($\mathrm{n, e}$, and $\mathrm{m_e}$ represent electron number density, elementary charge, and electron mass, respectively). The simulation involves 20 particles per cell (PPC) per species with a reduced proton-to-electron mass ratio of $\mathrm{m_p/m_e = 16}$, and the simulation time is measured in terms of $\mathrm{1/\omega_{p,e}}$. In PiC simulations of relativistic plasma, the use of a reduced $\mathrm{m_p/m_e}$ is a common practice to enhance computational efficiency and numerical stability. The large mass disparity between electrons and protons can lead to significant timestep discrepancies, where electrons require much smaller time steps than protons to accurately capture their dynamics. If we need to take the ion contribution into account, we should adjust the plasma frequency as $\mathrm{\omega_{p,e}^2 \rightarrow \omega_{p,e}^2(1 + m_e/m_p)}$. However, this modification is practically insignificant when dealing with the real $\mathrm{m_p/m_e}$ \citep[e.g.,][]{2013_Grismayer_second}. As we aim to study the underlying physics at electron scales and conduct analyses of simulation results covering time scales relevant for protons, utilizing the actual mass ratio is beyond our computational capabilities. Consequently, we opt for a reduced proton-to-electron mass ratio. PiC simulations have been employed to study magnetic field generation and particle acceleration within electron-proton plasmas in relativistic shear flows, considering both actual \citep[e.g.,][]{2014_Alves} and reduced \citep[e.g.,][]{2013_Grismayer} mass ratios. We set the speed of light as $\mathrm{c = 0.45 \Delta x/\Delta t}$, where $\mathrm{\Delta x}$ and $\mathrm{\Delta t}$ are the spatial and temporal resolutions, respectively. A correction factor for the speed of light is introduced, which causes electromagnetic waves to travel a little bit faster than the maximum speed of the particles. This helps ultra-relativistic flows to avoid the numerical Cerenkov instability, enabling stable and accurate simulations \citep[e.g.,][]{2007_anatoly, Sironi_2009}. 

The simulations are carried out in the equal Lorentz factor frame of reference (ELF) with an initial bulk Lorentz factor of $\Gamma_{\rm sp} = \Gamma_{\rm sh} = 15$. The right-moving spine plasma occupies the central 50\% of the numerical Y-grid, while the left-moving sheath plasma occupies the lower and upper quarters of the Y-grid. The shear interfaces are located at $\mathrm{Y = 512}$ and $\mathrm{Y = 1536}$ of the simulation box. Table \ref{table:tab_sim_parameter} summarizes the simulation parameters and their respective values used in this study.
\begin{figure}[H]
\includegraphics[width = \linewidth]{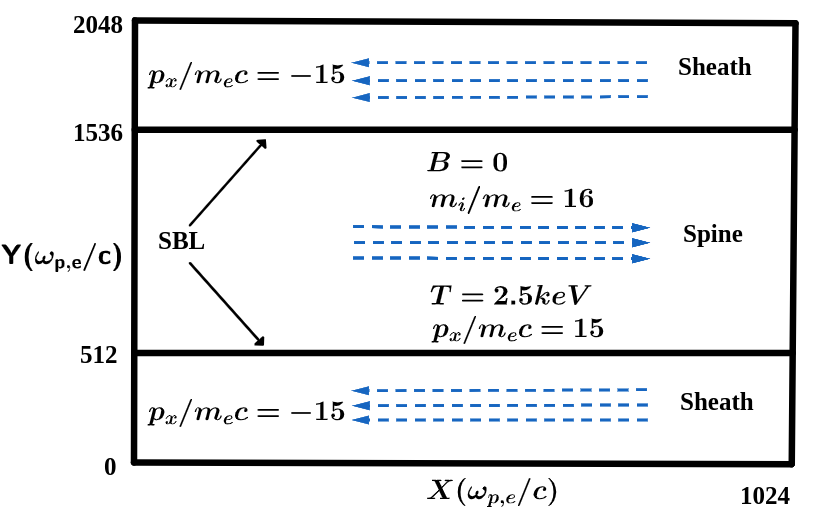}
\caption{2D simulation setup involving an electron-ion plasma with an initially unmagnetized shear flow: The plasma is composed of a central region in the Y-grid where right-moving plasma (spine) is located, where the top 25\% and bottom 25\% of the Y-grid are occupied by left-moving plasmas (sheath). The spine and sheath move in opposite directions with equal and opposite x-momenta $\mathrm{p_x/mc = \pm 15}$ in the ELF.}
\label{fig:setup}
\end{figure}
 \begin{deluxetable}{lccr}\label{table:tab_sim_parameter} 
\tablecaption{Table of simulation parameters} 
\tablehead{ 
    \colhead{Simulation parameters} &
    \colhead{Values} 
}
\startdata 
   PPC & 20   \\
   Plasma skin depth ($\mathrm{d_e}$) & 10 $\mathrm{\Delta x}$ \\
   $\mathrm{m_i/m_e}$ & 16 \\
   speed of light (c) & 0.45 $\mathrm{\Delta x/\Delta t}$ \\
   Correction factor for c & 1.025 \\
    $\mathrm{L_x}$ & 1024 $\mathrm{\Delta x}$\\
    $\mathrm{L_y}$ & 2048 $\mathrm{\Delta x}$\\
    Plasma density ($\mathrm{n_0}$) & PPC $\mathrm{\Delta x^{-3}}$ \\
    Plasma frequency ($\omega_{p,e}$) & 0.045$\mathrm{\Delta t^{-1}}$ \\
    Spatial resolution ($\mathrm{\Delta x}$) & 0.1 $\mathrm{d_e}$ \\
    Temporal resolution ($\mathrm{\Delta t}$) &  $0.045 \mathrm{\omega_{p,e}^{-1}}$ \\
\enddata
\end{deluxetable}
\FloatBarrier

\bigskip
\section{Inverse Compton scattering of External Blackbody Photons: Theory \label{sec: ICS-theory}}
In this section, we give a theoretical outline of our paper in which we derive the equations describing the inverse-Compton scattering of external photons. 

\bigskip
\subsection{Angle-averaged Inverse-Compton Emission in the Thomson Regime}
We consider inverse Compton cooling of accelerated electrons at SBLs in a thermal blackbody external photon field of varying temperatures with a characteristic frequency of $\mathrm{h\nu = 2.7 K_B T}$, where $\rm K_B$ and T are Boltzmann constant and radiation temperature, respectively. We first consider inverse-Compton cooling in the Thomson regime in which the scattering cross section is independent of the seed photon energy and the scattered photons are beamed along the direction of motion of the electrons. However, we use the full Klein-Nishina cross-section later while evaluating the angle-dependent radiation spectra (see section \ref{sec:ang_dep}). The differential Compton cross-section for a relativistic electron \citep[e.g.,][]{2009herb.book.....D,2012rjag.book.....B} is approximated using a hard cut-off at the transition to the Klein-Nishina regime:
\begin{align}\label{eq:cross-sec_AA}
    \begin{split}
    \frac{d\sigma}{d\Omega d\epsilon_s} &= \sigma_T \delta(\epsilon_s - \gamma^2 \epsilon_0) \delta(\Omega_s-\Omega_e)~H(1-\epsilon_0 \gamma) 
    \end{split}
   \end{align}
The inverse-Compton emissivity of a single electron in the Thomson limit can be expressed as
\begin{align}\label{eq:emissivity_AA}
   \begin{split}
    \operatorname{j_{\epsilon_0}^{head-on}(\gamma, \theta, \epsilon_0)} &=  \sigma_Tm_ec^3\epsilon_s\int_{0}^{\infty}n_{ph}(\epsilon_0, \theta) \eqbreak[1] \delta(\epsilon_s-\epsilon_0\gamma^2)\delta(\Omega_s-\Omega_e)~d\epsilon_0,
    \end{split}
\end{align}
where $\mathrm{\sigma_T}$ is Thomson cross section, $\mathrm{\epsilon_0 = h\nu/m_ec^2}$ and ${\rm \epsilon_s = h\rm \nu_s/\rm {m_ec^2}}$ refer to the initial and scattered photon energies, $\mathrm{\nu}$ and $\mathrm{\nu_s}$ being photon frequencies before and after the scattering. $\mathrm{\gamma}$ is the Lorentz factor of the relativistic electron scattering the soft target photons. The term $\mathrm{\delta(\Omega_s-\Omega_e)}$ specifies that the scattered photon travels in the direction of the incoming electron. The Heaviside function $\rm{H}$ cuts the cross-section off at the transitions to the Klein-Nishina regime. $\mathrm{\theta = K_BT/m_ec^2}$ is the temperature of soft photon field normalized by electron rest mass energy, where c  and $\mathrm{m_e}$ refer to the speed of light and electronic mass respectively. The density of the soft photon field $\mathrm{n_{ph}}$ is given by
\begin{equation}
    n_{ph}(\epsilon_0) = K \frac{\epsilon_0^2}{\exp(\frac{\epsilon_0}{\theta})-1}
    \label{eq:ph_dens}
\end{equation}
where, $\mathrm{K = 8\pi/\lambda_C^3}$ and $\mathrm{\lambda_C}$ is Compton wavelength. Equation \ref{eq:emissivity_AA}, when using a Dirac-delta approximation for the scattered photon energy, results in
\begin{align}\label{eq:emissivity_AA1}
    \operatorname{j_{\epsilon_s}^{head-on}(\gamma, \theta)} &= \sigma_Tm_ec^3K\delta(\Omega_s-\Omega_e) \frac{1}{\gamma^6} \frac{\epsilon_s^3}{\exp(\frac{\epsilon_s}{\gamma^2\theta})-1}.
\end{align}
The next step involves calculating the rate at which energy is lost by the electron as a result of inverse Compton cooling. To calculate this, we integrate Equation \ref{eq:emissivity_AA1} over the range of scattered photon energy ($\epsilon_s$). This integration yields the standard expression for the inverse Compton cooling rate of high-energy electrons in the Thomson regime within the sheath rest frame, where the radiation field is isotropic.
\begin{align}\label{eq:cooling-term_AA}
  \frac{d\gamma}{dt} &= \frac{\pi^4}{15} \sigma_TcK\gamma^2\theta^4
\end{align}
The radiation cooling term for inverse Compton scattering of relativistic electrons in the angle-integrated blackbody photon field in the Thomson regime in the spine frame is
\begin{align}\label{eq:cooling_term_AA_sp}
  \frac{d\gamma}{dt} &= \Gamma_{rel}^2\frac{\pi^4}{15} \sigma_TcK\gamma^2\theta^4,
\end{align}
where $\mathrm{\Gamma_{rel} = 2\Gamma^2 +1}$ is the relative bulk Lorentz factor between the spine and the sheath.
\bigskip
\subsection{Angle-dependent Radiation Spectra and Electron Cooling Rates}\label{sec:ang_dep}
In the evaluation of angle-dependent radiation spectra, we adopt a head-on approximation for the Compton cross section \citep[e.g.,][]{2009herb.book.....D,2012rjag.book.....B}, where all photons travel in the direction of the incoming electron. This allows us to simplify the Compton cross-section as follows:
\begin{equation}
   \frac{d\sigma}{d\Omega_s d\epsilon_s} = \delta (\Omega_s - \Omega_e) \frac{d\sigma}{d\epsilon_s}
\end{equation}
The differential Compton cross section can be integrated over $\Omega_s$ to get 
\begin{align}\label{eq:diff_cross}
    \frac{d\sigma_C}{d\epsilon_s} &= \frac{\pi r_e^2}{\gamma \epsilon^{'}} ~\bigg(y+\frac{1}{y}-\frac{2\epsilon_s}{\gamma \epsilon^{'}y}+ \big(\frac{\epsilon_s}{\gamma \epsilon^{'}y}\big)^2 \bigg) \\ \notag
    & \quad H\big( \epsilon_s~ ; ~ \frac{\epsilon^{'}}{2\gamma}~ ,~ \frac{2\gamma \epsilon^{'}}{1+2\epsilon^{'}} \big),
\end{align}
where $\epsilon^{'} = \gamma \epsilon_0 (1 - \beta \cos \psi)$ is the photon energy in electron rest frame, and $y = 1 - (\epsilon_s/\gamma)$. The emissivity of the electron is obtained by using the Compton cross section in equation (\ref{eq:diff_cross}) as:
\setlength{\abovedisplayskip}{4pt}
\setlength{\belowdisplayskip}{4pt}
\begin{equation}\label{eq:emissivity_AD}
\begin{aligned}
   \operatorname{j_C^{head-on}(\epsilon_s, \epsilon^{'}, \gamma)} &= \frac{3 m_e c^3 \sigma_T \epsilon_s (1-\beta\mu)}{8 \gamma} \int_0^\infty \frac{1}{\epsilon^{'}} \bigg\{y+\frac{1}{y} \\
   &\quad -\frac{2\epsilon_s}{\gamma \epsilon^{'}y} + \bigg(\frac{\epsilon_s}{\gamma \epsilon^{'}y}\bigg)^2\bigg\}~n_{ph}(\epsilon_0, \theta) \\
   &\quad H\bigg( \epsilon_s~ ; ~ \frac{\epsilon^{'}}{2\gamma}~ ,~ \frac{2\gamma \epsilon^{'}}{1+2\epsilon^{'}} \bigg)
\end{aligned}
\end{equation}
\begin{figure}[hbt!]
    \centering
\includegraphics[width=\columnwidth]{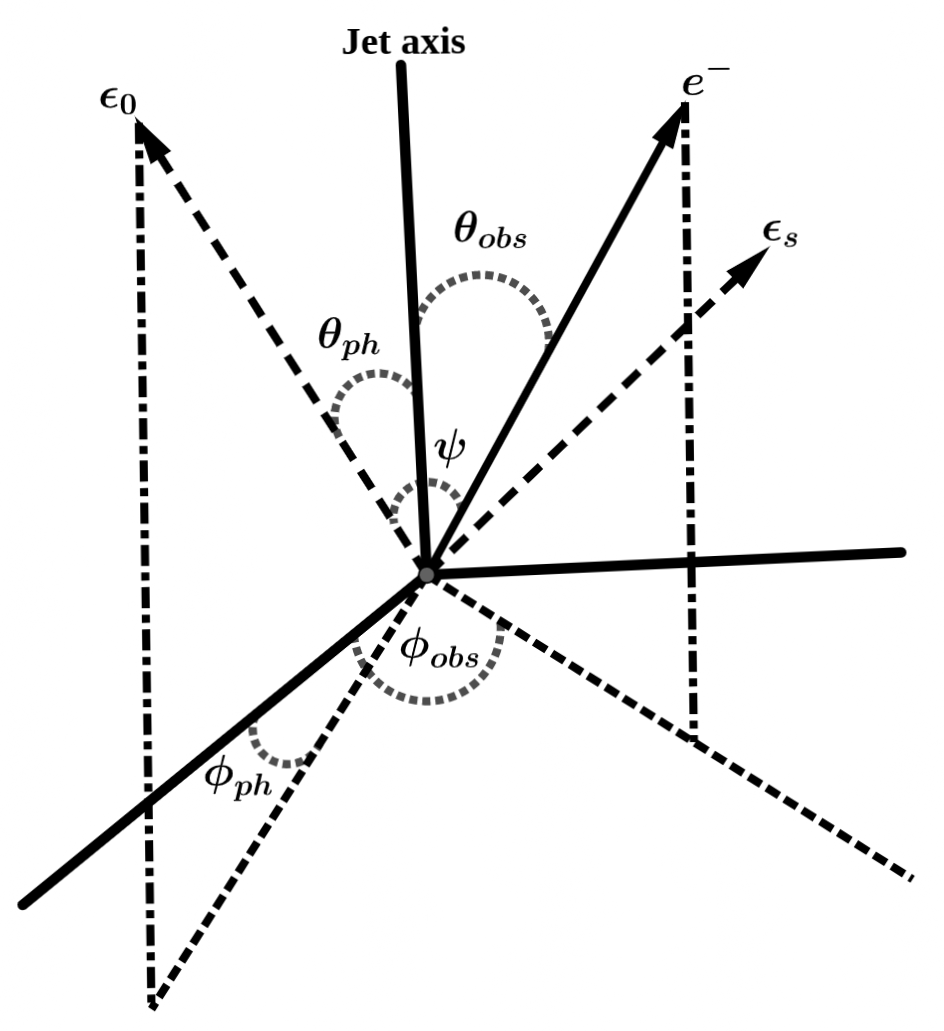}
\caption{Various angles in spherical geometry depicting Compton upscattering of photons with energy $\epsilon_0$ to $\epsilon_s$ off electrons with energies $\gamma$.}
\label{fig:ang_geometry}
\end{figure}

The blackbody radiation spectrum is strongly peaked at a mean photon energy of $\mathrm{<\epsilon_{0}(\theta)> = 2.7 \theta}$. Hence, a monochromatic $\delta$-function approximation can further simplify the spectral calculation to sufficient accuracy for a first exploration of the spectral and beaming patterns of such Compton emission \citep[e.g.,][]{2009herb.book.....D}. Thus, the blackbody photon density can be approximated as, $\mathrm{n_{ph} (\epsilon_0, \theta) = n_{ph}(\theta) \delta (\epsilon_0 - 2.7 \theta) = 2.4 K \theta^3}\delta (\epsilon_0 - 2.7 \theta)$. With this simplification, the Compton emission coefficient in Equation (\ref{eq:emissivity_AD}) can be expressed as:
\begin{align}\label{eq:emissivity_AD1}
    \operatorname{j(\epsilon_s, \theta, \gamma)} &= \frac{0.3 m_e c^3\sigma_T K \epsilon_s \theta^2}{\gamma^2} \bigg[y+\frac{1}{y}-\frac{2\epsilon_s}{2.7 y\gamma^2\theta (1-\beta \mu)} \\ \notag
    & \quad + \frac{\epsilon_s^2}{(2.7)^2\gamma^4 \theta^2 (1-\beta \mu)^2y^2} \bigg] \\ \notag
    & \quad H\bigg( \epsilon_s ; \frac{2.7 \theta (1-\beta \mu)}{2} \notag
     ,\frac{5.4\gamma^2 (1-\beta \mu)\theta}{1+5.4\gamma \theta (1-\beta \mu)} \bigg)
\end{align}
Equation (\ref{eq:emissivity_AD1}) needs to be integrated over scattered photon energies, $\epsilon_s$ within the limits constrained by the Heaviside function to get the Compton cooling term. Equation (\ref{eq:emissivity_AD1}) is the general expression for Compton emissivity that is applicable in both the Thomson and
\begin{figure*}[hbt!]
\centering
\includegraphics[width=0.965\linewidth]{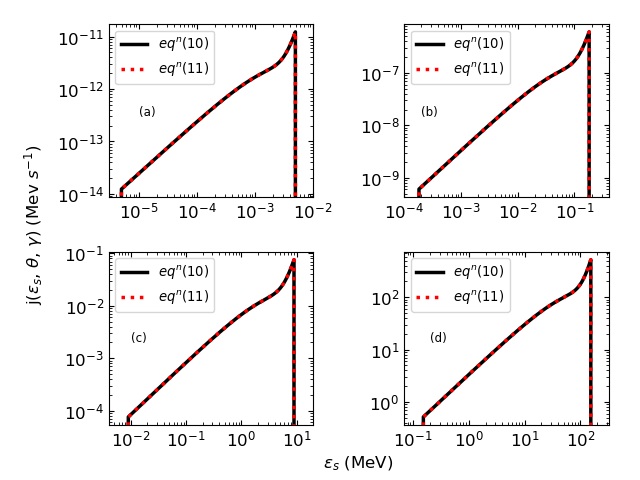}
\caption{Comparison between equations (\ref{eq:emissivity_AD1}) and (\ref{eq:emissivity_AD2}), for $\mathrm{\gamma = 10^3}$ and $\mathrm{\cos \psi = -1}$: The top panel shows the Compton emissivity due to a single electron vs. scattered photon energy for $\mathrm{\theta = 4.58 \times 10^{-10}}$ (left) and $\mathrm{\theta = 10^{-8}}$ (right), respectively whereas the lower panel shows the same for $\mathrm{\theta = 10^{-6}}$ (left) and $\mathrm{\theta = 10^{-5}}$ (right), respectively.}
\label{fig:emisivty_eps_s}
\end{figure*}
\FloatBarrier
\noindent Klein-Nishina regimes. The term $\mathrm{y + 1/y}$ can be rewritten as $\mathrm{2 + \frac{\epsilon_s^2}{\gamma (\gamma - \epsilon_s)}}$. In the Thomson regime, when $\epsilon^{'} <<1$, $\gamma >> \epsilon_s$ and $\gamma - \epsilon_s \cong \gamma$. In this limit, equation (\ref{eq:emissivity_AD1}) can be simplified by binomially expanding the terms containing ($\gamma - \epsilon_s$) as
\begin{align} \label{eq:emissivity_AD2}
\operatorname{j(\epsilon_s, \theta, \gamma)} &= \frac{0.3c\sigma_T K \theta^2}{\gamma^2} \Bigg[2\epsilon_s +   \epsilon_s \bigg\{ \bigg(\frac{\epsilon_s}{\gamma}\bigg)^2 + \bigg(\frac{\epsilon_s}{\gamma}\bigg)^3 \\ \notag
& \quad + .... \bigg\} -\frac{2}{2.7\theta(1-\beta\mu)} \bigg\{ \bigg(\frac{\epsilon_s}{\gamma}\bigg)^2 
 +  \bigg(\frac{\epsilon_s}{\gamma}\bigg)^3 \\ \notag
 & \quad  + ....\bigg\}  + \frac{1}{2.7^2 \theta^2 (1-\beta\mu)^2} \frac{1}{\gamma} \bigg(\frac{\epsilon_s}{\gamma} \bigg)^3 \bigg\{ 1 +  \\ \nonumber 
 & \quad 2\bigg(\frac{\epsilon_s}{\gamma}\bigg) + 3 \bigg( \frac{\epsilon_s}{\gamma} \bigg)^2 + 4 \bigg(\frac{\epsilon_s}{\gamma} \bigg)^3 + ....  \bigg\}\Bigg] \\ \notag
 & \quad H\bigg( \epsilon_s~ ; ~  \epsilon_{\mathrm{s_{min}}} ,\epsilon_{\mathrm{s_{max}}} \bigg),
\end{align}
\noindent where $\mathrm{\epsilon_{s_{max}} = 5.4\gamma^2 (1-\beta \mu)\theta/(1+5.4\gamma \theta (1-\beta \mu))}$ and $\mathrm{\epsilon_{s_{min}} = 2.7 \theta (1-\beta \mu)/2}$ are the maximum and minimum energies of the upscattered photons. Equations (\ref{eq:emissivity_AD1}) and (\ref{eq:emissivity_AD2}) were compared to assess the impact of the binomial approximation. Figure \ref{fig:emisivty_eps_s} indicates that the approximation has no significant effect on the result for electrons with Lorentz factors of $\gamma = 10^3$, even when scattering UV target photons with $\mathrm{\mu = -1}$. However, when dealing with higher values of $\gamma$ and $\epsilon_0$, it is essential to consider the higher order terms of the binomial expansion that contain ($\mathrm{\gamma-\epsilon_s}$) to avoid significant deviations from the full results. Due to the presence of the Heaviside function in the emissivity equations, (\ref{eq:emissivity_AD1}) and (\ref{eq:emissivity_AD2}), a sudden drop in emissivity occurs beyond the scattered photon energy, $\mathrm{\epsilon_s = \epsilon_{s_{max}}}$.  By employing the aforementioned approximations, the integration of equation (\ref{eq:emissivity_AD2}) over $\mathrm{\epsilon_s}$ becomes feasible, considering the limits imposed by the Heaviside function. 
\begin{figure*}[hbt!]
\centering
\includegraphics[width=\linewidth]{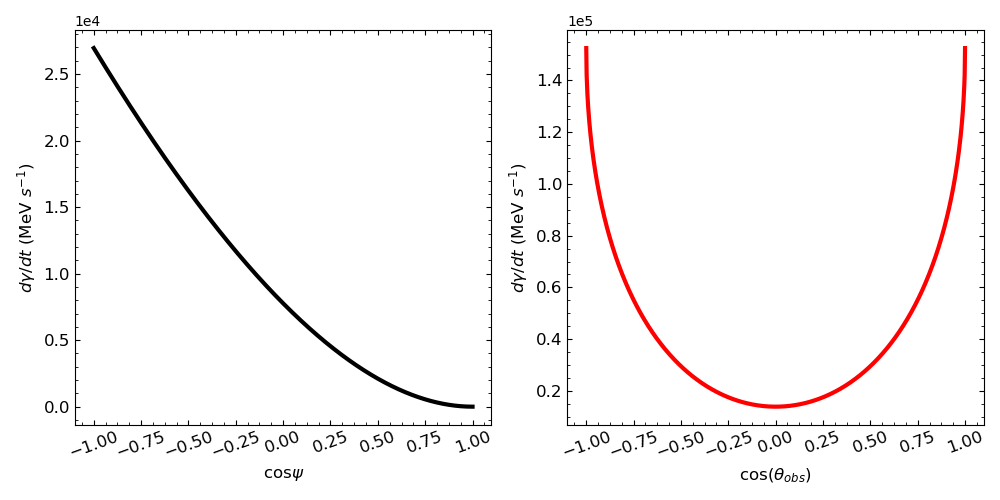}
\caption{Variation of cooling rate with the cosine of collision angle ($\mathrm{\cos\psi}$) and observer's angle ($\mathrm{\cos(\theta_{\text{obs}}}$)) in the ELF at a Lorentz factor ($\mathrm{\gamma}$) of $\mathrm{10^3}$ and radiation temperature ($\mathrm{\theta}$) of $\mathrm{10^{-5}}$: the plot on the right depicts the inverse cooling term, obtained by integrating equation (\ref{eq:cooling_term_AD}) over the photon distribution angle ($\mathrm{\mu_{\text{ph}}}$). The figure illustrates that the maximum cooling occurs during head-on collisions between electrons and photons, or when electrons, after scattering, travel parallel to the jet axis. Conversely, the minimum cooling is observed for tail-on collisions or when electrons travel perpendicular to the jet axis.}
\label{fig:cooling_rate_vs_mu}
\end{figure*}
\FloatBarrier
The cooling term associated with inverse Compton scattering in the sheath frame, taking into account an angle-dependent photon field, can be expressed as\\
\begin{align}\label{eq:cooling_term_AD}
\frac{d \gamma}{dt}(\mu, \gamma, \theta) &= A \Bigg[\big(\epsilon_{s_{\mathrm{max}}}^2-\epsilon_{s_{\mathrm{min}}}^2\big) + \frac{\epsilon_{s_{\mathrm{max}}}^4 - \epsilon_{s_{\mathrm{min}}}^4}{4\gamma^2} \\ \nonumber 
&\quad+ \frac{\epsilon_{s_{\mathrm{max}}}^5 - \epsilon_{s_{\mathrm{min}}}^5}{5\gamma^3} - \frac{2}{B} \bigg\{\frac{(\epsilon_{s_{\mathrm{max}}}^3 - \epsilon_{s_{\mathrm{min}}}^3)}{3\gamma^2} \\ \nonumber 
&\quad+ \frac{\epsilon_{s_{\mathrm{max}}}^4 - \epsilon_{s_{\mathrm{min}}}^4}{4\gamma^3}  \bigg\} + \frac{1}{\gamma^4B^2} \bigg\{\frac{\epsilon_{s_{\mathrm{max}}}^4 - \epsilon_{s_{\mathrm{min}}}^4}{4} \\ \nonumber
&\quad + \frac{2\big(\epsilon_{s_{\mathrm{max}}}^5 - \epsilon_{s_{\mathrm{min}}}^5 \big)}{5\gamma} + \frac{\epsilon_{s_{\mathrm{max}}}^6 - \epsilon_{s_{\mathrm{min}}}^6}{2\gamma^2}\\ \nonumber 
&\quad + \frac{4\big(\epsilon_{s_{\mathrm{max}}}^7 - \epsilon_{s_{\mathrm{min}}}^7 \big)}{7\gamma^3}\bigg\} \Bigg],
\end{align}

where $\mathrm{A \equiv 0.3~c~\sigma_T K \theta^2/\gamma^2}$ and $\mathrm{B \equiv 2.7\theta(1-\beta\mu)}$. To calculate the cooling rate in the spine frame, a factor $\mathrm{\Gamma_{rel}^2}$ needs to be added to the r.h.s. The angular distribution of inverse Compton scattering is described by the expression $\mu = \mu_{obs} \mu_{ph} + \sqrt{1 - \mu_{obs}^2} \sqrt{1 - \mu_{ph}^2} \cos(\phi_{obs} - \phi_{ph})$. Here, $\mathrm{\mu_{obs}}$ and $\mathrm{\mu_{ph}}$ correspond to the cosines of the angles $\mathrm{\theta_{obs}}$ and $\mathrm{\theta_{ph}}$, respectively. These angles indicate the directions of the electron and photon relative to the jet axis. By assuming the azimuthal symmetry of the system and performing the numerical integration of equation (\ref{eq:cooling_term_AD}) over $\mathrm{\mu_{ph}}$, the electron cooling rate can be obtained as a function of $\mathrm{\mu_{obs}}$. Figure \ref{fig:cooling_rate_vs_mu} illustrates the dependency of the cooling rate on the cosine of $\mathrm{\psi}$ and $\mathrm{\theta_{obs}}$, highlighting the variation of the electron cooling rate with respect to these angular parameters. The plots are generated in the ELF at electron Lorentz factor of $\mathrm{\gamma = 10^3}$ and the radiation temperature of $\mathrm{\theta = 10^{-5}}$.
\bigskip


\section{Results \label{sec:results}}
This section presents the outcomes of our study concerning self-generated electric and magnetic fields in SBLs of relativistic jets, along with the effects of particle anisotropy, radiative output, and IC cooling on particle dynamics obtained through PiC simulations.

\subsection{Self-generated magnetic and electric fields \label{subsec:e_m_fields}}

\begin{figure*}[!htb]
  \centering
  \begin{tabular}{@{}c@{}c@{}}
  \centering
    \subfigimg[width=0.5\linewidth,height=0.245\paperheight, keepaspectratio, pos=ur, vsep=1cm, hsep=2.5cm]{\textbf{(a)}}{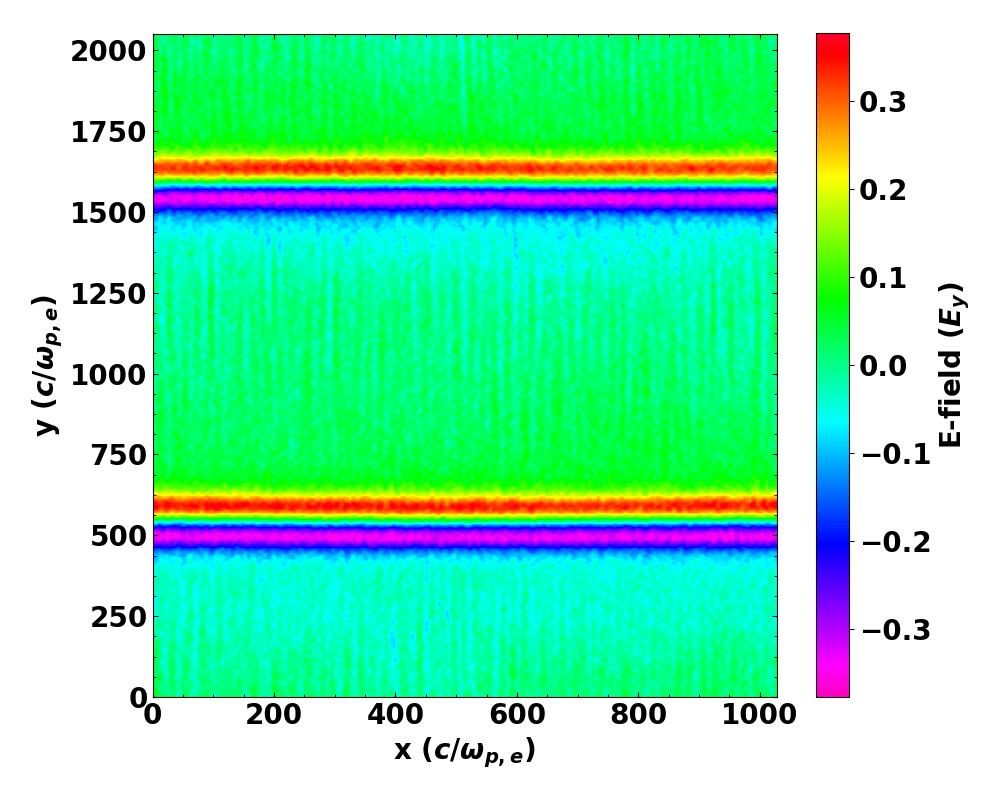} &
    \subfigimg[width=0.5\linewidth,height=0.245\paperheight, keepaspectratio,, pos=ur, vsep=1cm, hsep=2.5cm]{\textbf{(b)}}{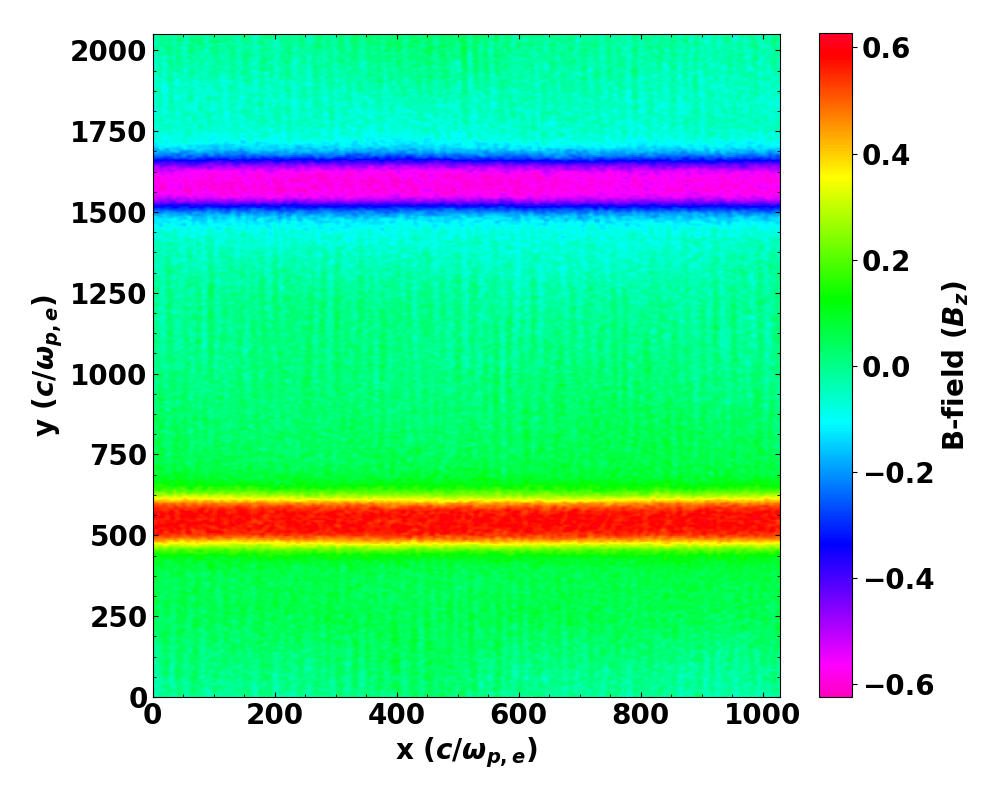} \\
    \subfigimg[width=0.5\linewidth,height=0.245\paperheight, keepaspectratio, pos=ur, vsep=1cm, hsep=2.5cm]{\textbf{(c)}}{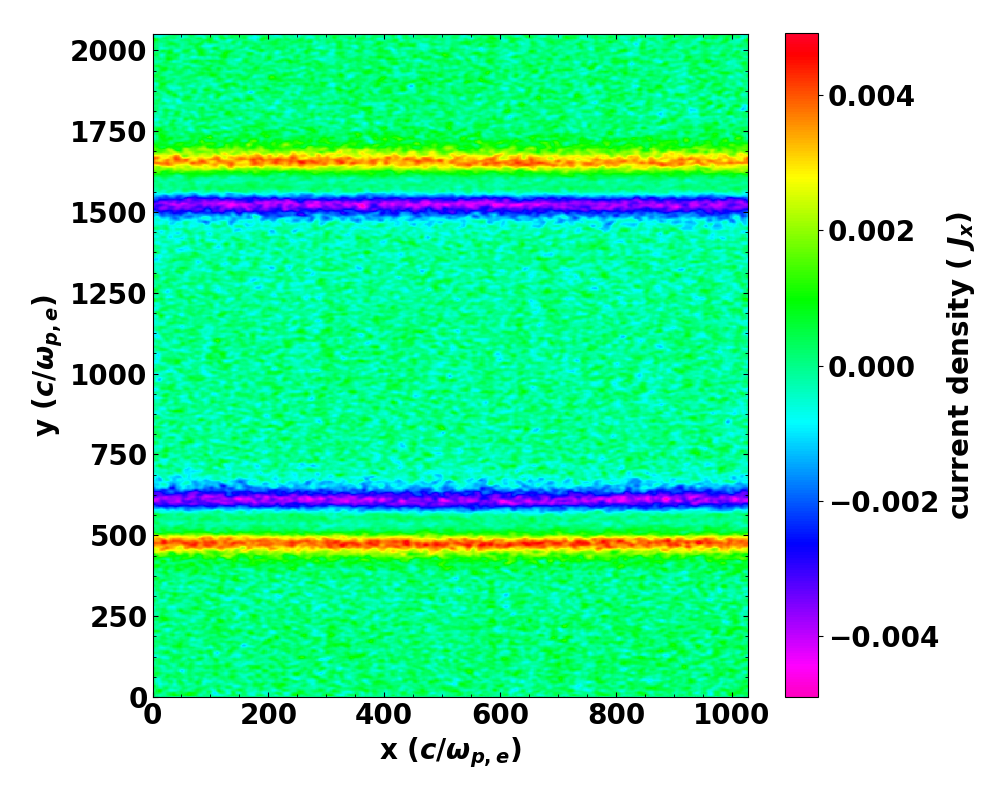} &
    \subfigimg[width=0.5\linewidth,height=0.245\paperheight, keepaspectratio, pos=ur, vsep=1cm, hsep=2.5cm]{\textbf{(d)}}{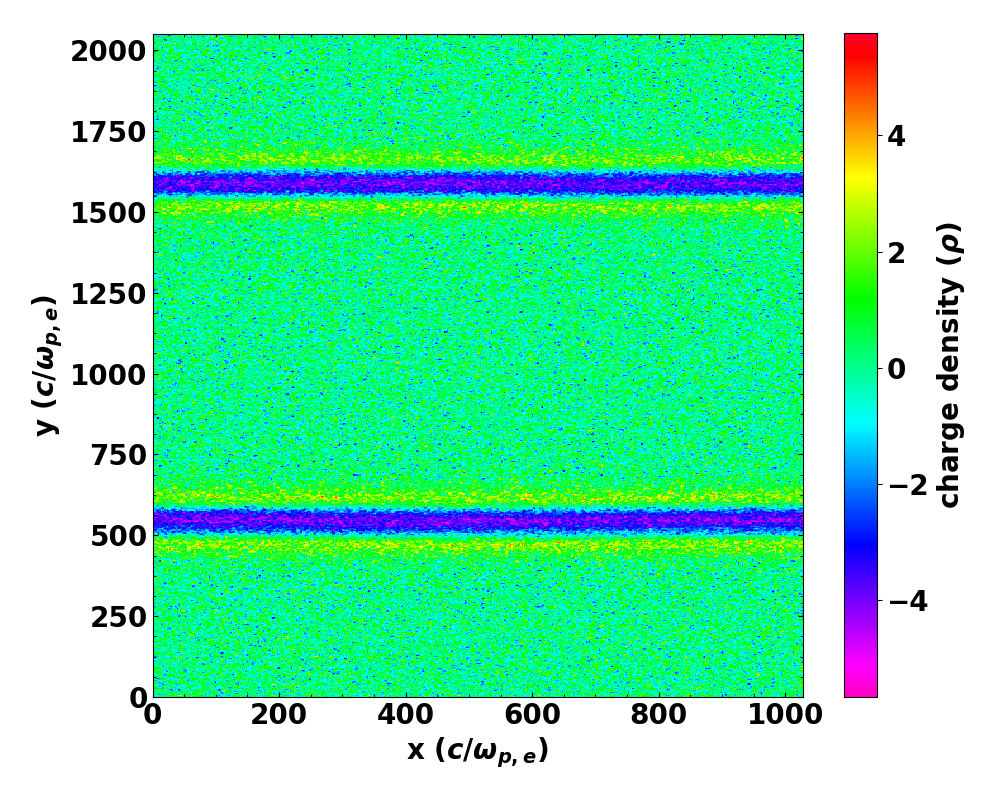} \\
    \subfigimg[width=0.5\linewidth,height=0.245\paperheight, keepaspectratio, pos=ur, vsep=1cm, hsep=2.5cm]{\textbf{(e)}}{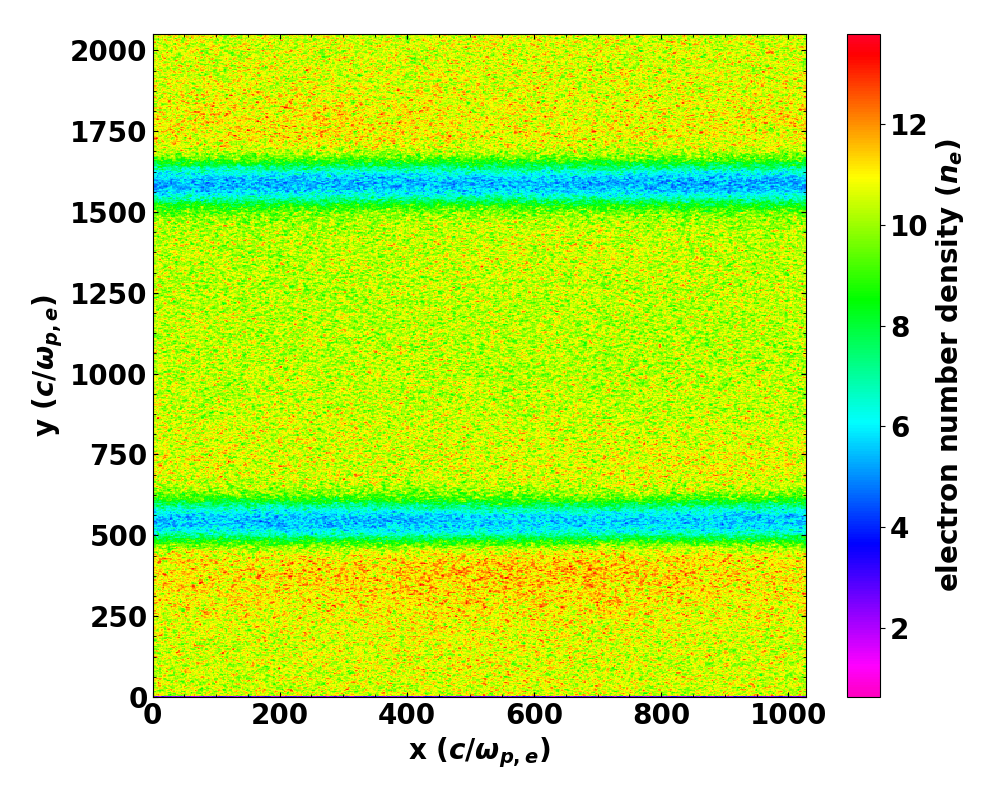} &
    \subfigimg[width=0.5\linewidth,height=0.245\paperheight, keepaspectratio, pos=ur, vsep=1cm, hsep=2.5cm]{\textbf{(f)}}{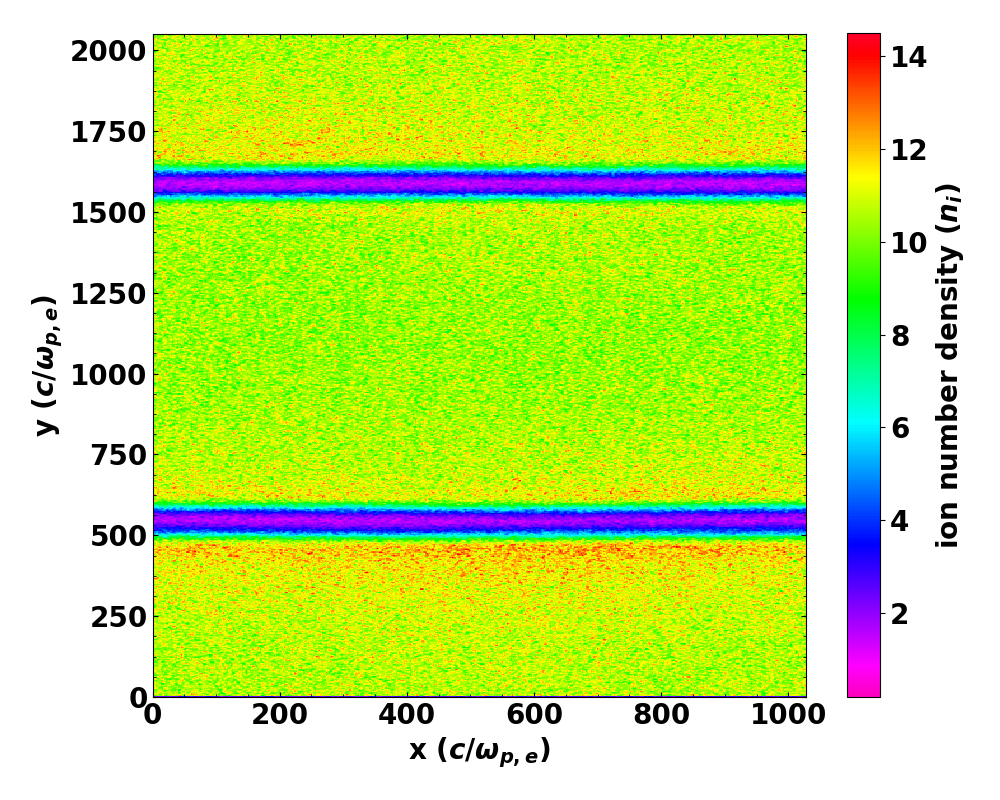}
  \end{tabular}
  \caption{First row: Figure (a) illustrates the xy-cut of the self-generated electric field and Figure (b) presents the magnetic field in SBLs, with the red and magenta colors representing opposite polarities. Second row: Figure (c) depicts the current distribution ($\mathrm{J_x}$). The ions dominate the stronger outer current sheet, while the weaker inner current sheet is dominated by electrons. Figure (d) shows the total charge density ($\mathrm{\rho}$). Third row: Figures (e) and (f) respectively show the density distribution of electrons and ions. Notably, a discernible depletion in particle density is observed in the vicinity of the boundary layers. All figures correspond to $\mathrm{t = 3000\omega_{p,e}^{-1}}$. The units displayed are arbitrary.}
    \label{fig:E-B_J_n}
\end{figure*}

Plasma instabilities like the Weibel instability self-generate a magnetic field along the z-direction (perpendicular to the jet axis). The self-generated electromagnetic fields in SBLs create turbulence in the spine-sheath interface which eventually leads to particle acceleration up to TeV energy in relativistic jets. Figures \ref{fig:E-B_J_n}(a), \ref{fig:E-B_J_n}(b), and \ref{fig:E-B_J_n}(c) show the dominant components of electric and magnetic fields and the current density produced from the PiC simulation.
\FloatBarrier
\begin{figure}[!htb]
\centering
\includegraphics[height = 8cm,width=\columnwidth]{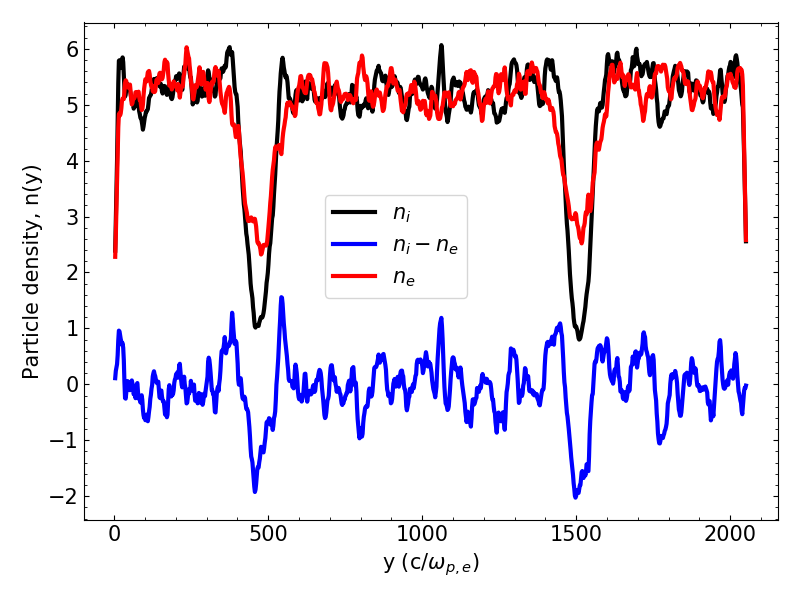}
\caption{The electron number densities as a function of y at $t = 3000 \omega_{p,e}^{-1}$: while the ions are fully expelled from the shear interface, the electrons create a layer near the interface, resulting in the formation of a triple layer due to charge separation.}
\label{fig:density_prof}
\end{figure}
\FloatBarrier
Figure \ref{fig:E-B_J_n}(d) illustrates the net charge distribution in the SBLs, showing a notable concentration of negatively charged particles (electrons) within these regions. Positively charged particles (ions), on the other hand, tend to have a slightly more dispersed distribution and are found slightly further from the SBLs, thereby forming a triple-layer structure. Figures \ref{fig:E-B_J_n}(e) and \ref{fig:E-B_J_n}(f) present the spatial distribution of electron and ion number densities, respectively. These figures highlight higher electron density within the SBLs compared to the ion density. This disparity in densities is further evident in the density profiles of both species (Figure \ref{fig:density_prof}). The density profiles of ions, electrons, and net charge as a function of the y-coordinate are demonstrated in Figure \ref{fig:density_prof}. Due to the different velocities and densities of ions and electrons, the current density along the x-direction ($\mathrm{J_x}$) (as depicted in Figure \ref{fig:E-B_J_n}(c)) produces the uniform magnetic field ($\mathrm{B_z}$) along the shear interface within the shear flows. Because the electron gyroradius is smaller than that of ions, ions are more displaced than electrons, leading to their expulsion from the shear interface due to the extra magnetic pressure, resulting in the formation of a density trough. This process induces charge separation and the formation of a triple layer, as illustrated in Figure \ref{fig:density_prof}. Ion-dominated current sheets are formed farther from the shear interface, while electron-dominated current sheets are formed near it, with an electric field ($\mathrm{E_y}$) pointing from the ion-dominated current sheet to the electron-dominated current sheet. Ultimately, the electrons undergo efficient acceleration driven by the cross fields ($\mathrm{E \times B}$) in the x-direction, which aligns with the jet-propagation direction. As the ions are more massive than electrons, ions undergo free streaming while electrons behave like fluid \citep[e.g.,][]{gruzinov2008grb, 2017_Liang}.
 The kinetic KHI plays a vital role in initially cold electrons. However, the electrons attain a finite transverse momentum either by plasma instabilities or due to the finite temperature achieved by electrons. Eventually, electrons with finite transverse momentum cross over to the sheath moving in the opposite direction, but heavy ions do not. This mechanism, referred to as the electron counter-current instability (ECCI) \citep[e.g.,][]{2017_Liang}, dominates, leading to the formation of opposite current layers on the two sides of the shear interface, creating a monopolar slab of magnetic field.
\bigskip
\subsection{Particle Anisotropy in SBLs \label{subsec:part_aniso}}
\begin{figure}[!htb]
\includegraphics[width=\columnwidth,height=7.5cm]{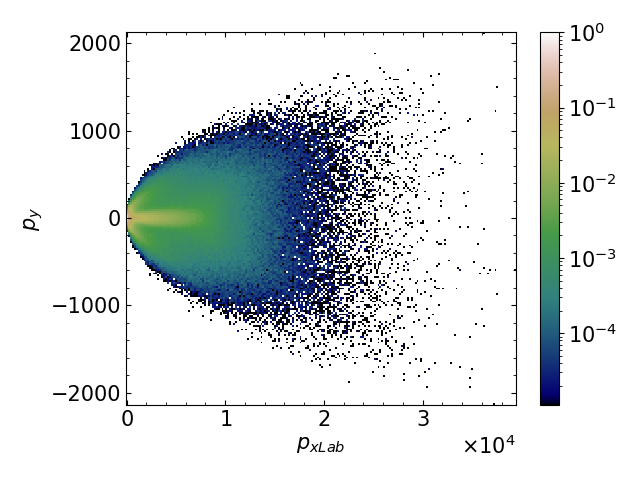}
\caption{Spine electrons' $\mathrm{p_y}$ versus $\mathrm{p_{xLab}}$ at time $\mathrm{t = 3000 \omega_{p,e}^{-1}}$: $\mathrm{p_x}$ are Lorentz boosted to the Laboratory frame by $\mathrm{\Gamma = 15}$. The figure illustrates that some of the spine electrons diffuse into the sheath region, leading to the deceleration of those spine electrons, represented by the low-energy arc-shaped electron population. The high-energy electron population towards the right-hand side of the figure corresponds to electrons that remain in the spine region without crossing over into the sheath region.}
\label{fig:py_vs_pxlab}
\end{figure}
\FloatBarrier
In order to illustrate the particle acceleration and radiation properties in the stationary (sheath) frame of the source, we Lorentz-transform the particle momenta and Compton emission into the sheath frame, utilizing a bulk Lorentz factor of $\mathrm{\Gamma = 15}$. The results demonstrate that spine electrons undergo significant acceleration,
\begin{figure}[!htb]
\includegraphics[height=6.5cm,width=\columnwidth]{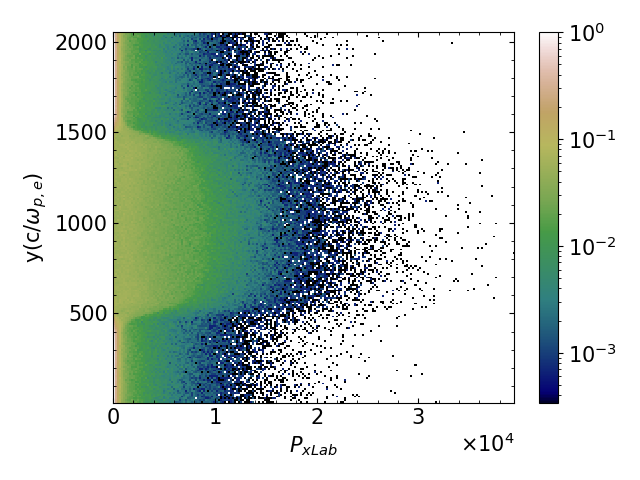}
\caption{$\mathrm{y}$ ($\mathrm{c/\omega_{p,e}}$) versus $\mathrm{p_{xLab}}$ of spine electrons at $\mathrm{t = 3000 \omega_{p,e}^{-1}}$: $\mathrm{p_x}$ are Lorentz boosted to the laboratory frame. As depicted in the figure, some of the spine electrons cross over through SBLs and enter the sheath region, where they experience deceleration. The spine electrons that remain strictly within the spine region undergo acceleration.}
\label{fig:y_vs_pxlab}
\end{figure}
\FloatBarrier
\begin{figure}[!htb]
\centering
\includegraphics[height = 8cm,width=\columnwidth]{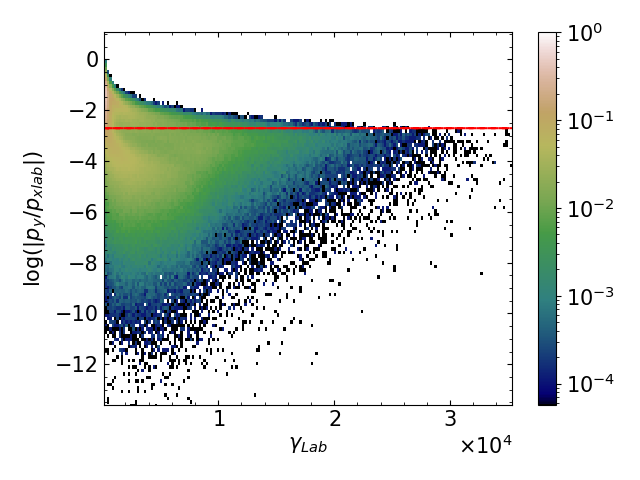}
\caption{The distribution of the tangent of the beam angle of spine electrons versus electron Lorentz factor in the laboratory frame at $\mathrm{t = 3000 \omega_{p,e}^{-1}}$: all high-energy spine electrons, which did not cross over to the sheath region, possess beam angles significantly smaller than 1/$\mathrm{\Gamma}$, as indicated by the red line. The figure further illustrates an anticorrelation between beam angle and electron energy.}
\label{fig:beam_ang_vs_gam_lab}
\end{figure}
\FloatBarrier
\noindent attaining high energies and exhibiting pronounced momentum anisotropy ($\mathrm{p_{xlab} >> p_y}$) in the laboratory frame (Figure \ref{fig:py_vs_pxlab}). A portion of the spine electrons diffuses into the sheath region, experiencing deceleration and giving rise to a low-energy population, as depicted at the left in Figures \ref{fig:py_vs_pxlab} and \ref{fig:y_vs_pxlab}. Moreover, Figure \ref{fig:y_vs_pxlab} illustrates the large particle momenta along the jet axis in the spine. Furthermore, we observe an anti-correlation between the beam angle $\mathrm{|p_y/p_{\text{xLab}}|}$ and electron energies, with the beam angle decreasing as electron energies increase (Figure \ref{fig:beam_ang_vs_gam_lab}).

Notably, the beam angles of all high-energy spine electrons, which remained in the spine region and did not transition to the sheath region, are significantly smaller than those resulting from Doppler boosting an isotropic particle distribution from the spine rest frame to the laboratory frame ($\mathrm{1/\Gamma}$), as depicted by the red dashed line in Figure \ref{fig:beam_ang_vs_gam_lab}.
\bigskip
\subsection{Numerical Cerenkov Instability}
The Numerical Cerenkov Instability (NCI) is a significant concern in PiC simulations that involve relativistic drifts. NCI is a computational artifact that arises in PiC simulations due to the discretization scheme employed. It is not a physical instability but a numerical phenomenon. In PiC simulations, electromagnetic fields and particle motions are discretized on a computational grid. When the discrete grid points interact with the chosen spatial and temporal resolutions, NCI occurs generating spurious wave modes. These modes can distort simulation results and potentially impact the accuracy of the physical processes under the study. The momentum plot of the magnetic field provides insight into the presence and effects of NCI.

To mitigate the NCI, the \texttt{TRISTAN-MP} code incorporates a range of specialized numerical schemes that help suppress the instability. It employs both time-centered and space-centered finite difference schemes to ensure second-order accuracy in both space and time. Time-centered schemes calculate values at time step midpoints for enhanced time integration precision, while space-centered schemes compute spatial derivatives at midpoint grid locations for improved spatial accuracy. Magnetic and Electric fields are stored within a 3D Yee mesh \citep[e.g.,][]{1966I_Yee}, effectively capturing their spatial distribution. A tri-linear interpolation function relates these fields to particle positions, ensuring accurate values for simulated particle positions. A three-point digital binomial filter \citep[e.g.,][]{106800,hockney2021computer} is applied along each spatial dimension to source terms, using weights of 0.25, 0.5, and 0.25 to mitigate NCI and other non-physical high-frequency field modes arising from finite difference calculations. The code
\begin{figure}[!htb]
  \centering
  \begin{tabular}{c}
    \subfigimg[width=\columnwidth, pos = ul, hsep = 2 cm, vsep = 1cm]{\color{white}\textbf{(a)}}{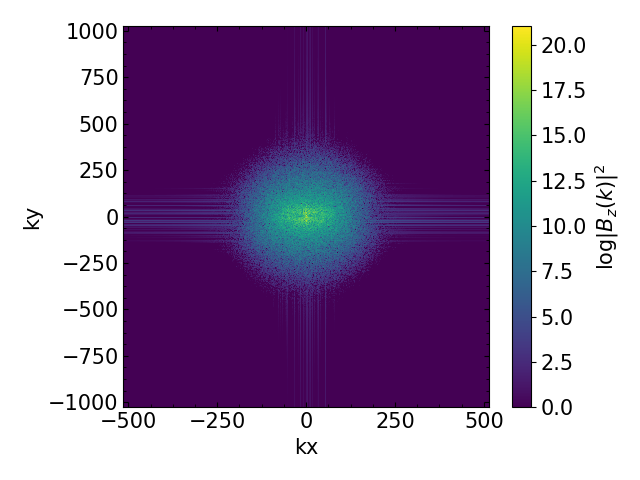} \\
    \subfigimg[width=\columnwidth, pos = ul, hsep = 2 cm, vsep = 1cm]{\color{white}\textbf{(b)}}{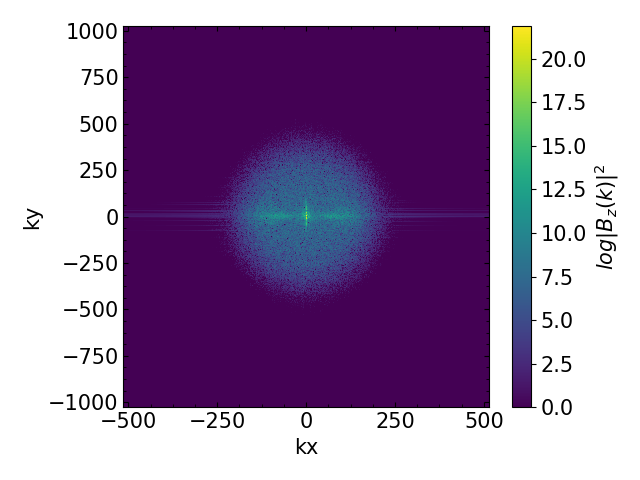} \\
  \end{tabular}
  \caption{2D color contour plots of the Fourier amplitude of $\mathrm{B_z}$ at two different sample times, (a) $\mathrm{t=200\omega_{p,e}^{-1}}$ and (b) $\mathrm{t=3000\omega_{p,e}^{-1}}$: It can be observed that the effects of NCI are minimal in the simulation, and they tend to gradually decrease as the simulation progresses.}
  \label{fig:NCI}
\end{figure}
\FloatBarrier
\noindent offers a choice between a second-order method and a fourth-order stencil approach \citep[e.g.,][]{2004_Greenwood}. We opt for the latter approach as it is aimed at minimizing NCI effects, particularly in relativistic scenarios. The code adheres to the Courant-Friedrichs-Lewy (CFL) condition, ensuring that the product of the timestep and the speed of light is smaller than the minimum cell size. In this context, the speed of light is set as $\rm{c = 0.45 \Delta x/\Delta t}$. This also serves to mitigate the effects of the NCI. As a result, the current simulations are able to contain the NCI within acceptable levels, leading to minimizing adverse effects on particle heating. The $\mathrm{kx-ky}$ plot in Figure (\ref{fig:NCI}) represents the distribution of the $\mathrm{z}$-component of the magnetic field in momentum space, with $\mathrm{kx}$ and $\mathrm{ky}$ representing the wave numbers in the x and y directions, respectively. The smooth and well-behaved distribution of the magnetic field in the $\mathrm{kx-ky}$ plot suggests a minimal effect of NCI indicating that the chosen parameters in the simulations adequately mitigate numerical artifacts.
\bigskip
\subsection{Electron Spectra and Effect of IC Cooling \label{subsec:elec_spec}}
\begin{figure}[!htb]
  \centering
  \begin{tabular}{c}
    \subfigimg[width=\linewidth, pos = ur, vsep = 1cm]{\color{black}\textbf{(a)}}{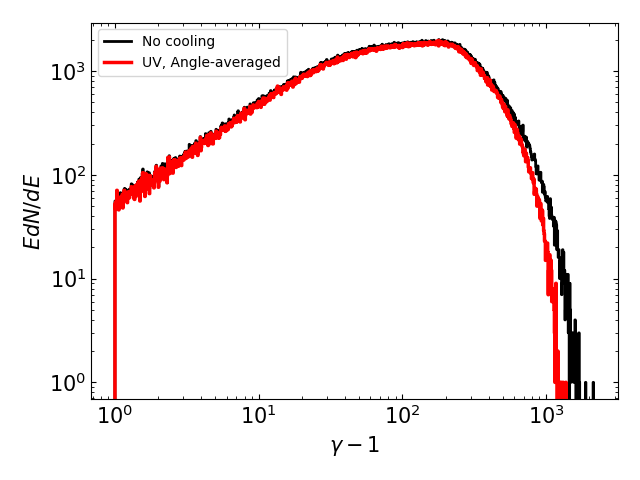} \\
    \subfigimg[width=\linewidth, pos = ur, vsep = 1cm]{\color{black}\textbf{(b)}}{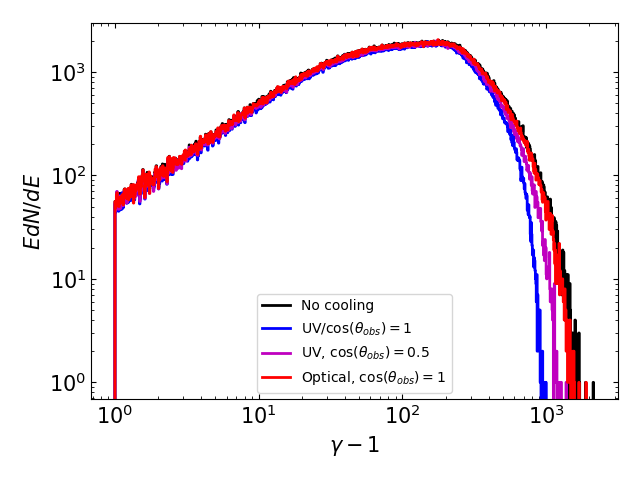} \\
  \end{tabular}
  \caption{Comparison of electron Spectra with and without inverse Compton cooling by blackbody photons: Figure (a) displays the impact of IC cooling due to angle-averaged UV photons, while Figure (b) exhibits the effect of IC cooling induced by angle-dependent UV and optical photons at various values of $\mathrm{\cos(\theta_{obs}}$).}
  \label{fig:elec_cooling}
\end{figure}
\FloatBarrier
This section focuses on electron acceleration with PiC simulations demonstrating the impact of IC cooling on particle spectra. Initially, the energy distribution of electrons follows a Maxwellian distribution. Acceleration occurs due to the electromagnetic field generated by the relative displacement of ions and electrons. The thermal electron bulk Lorentz factor reaches its peak when energy equipartition between the ions and electrons is attained \citep[e.g.,][]{Lyubarsky:2006ix, 2019}. Thus, the maximum electron Lorentz factor can be expressed as:
\begin{align*}
    \gamma_{max} \simeq \frac{1}{2}\frac{\gamma_i m_i}{m_e}
\end{align*}
The spectra illustrated in Figure \ref{fig:elec_cooling} are measured in the simulation frame, where a pure $e^-$-ion plasma shows no evidence of a power law. Initially, the electron spectra exhibit a Maxwellian distribution that evolves into a non-Maxwellian distribution, with a remnant of the original distribution. Ultimately, the system reaches a steady state with a peak around $\mathrm{\gamma \cong \Gamma m_i/m_e}$.

To investigate the impact of radiation cooling on the electron spectra, the cooling terms presented in equations \ref{eq:cooling_term_AA_sp} and \ref{eq:cooling_term_AD} can be incorporated into the electron subroutine of the PiC code. Figure \ref{fig:elec_cooling} illustrates the impact of Inverse Compton (IC) cooling on electron spectra induced by angle-averaged and angle-dependent photons. The strength of IC cooling depends on the radiation energy density and the Lorentz factor of the electrons. At lower electron energies, the cooling effect is negligible, while at higher energies, it significantly affects electron acceleration and the energy spectrum. Due to radiation drag, the high-energy end of the spectrum shifts towards lower energies (as seen in Fig. \ref{fig:elec_cooling}). Compared to IC cooling induced by UV photons, the effect of Compton drag for photons with temperatures $\theta \le 10^{-6}$ is subdominant, and thus does not appreciably affect particle dynamics and the electron energy spectrum.
\bigskip
\subsection{Self-consistent Radiation Spectra \label{subsec:rad_spec}}
We now discuss the radiation spectra resulting from Compton upscattering of different radiation backgrounds. Specifically, we consider the cosmic microwave background, infrared, optical, and ultraviolet photons with corresponding values of $\theta$ equal to $\mathrm{4.58 \times 10^{-10}}$, $\mathrm{10^{-8}}$, $\mathrm{10^{-6}}$, and $\mathrm{10^{-5}}$, respectively. 

\begin{figure}[!htb]
  \centering
  \begin{tabular}{c}
    \subfigimg[width=\linewidth, pos = ur, vsep = 1 cm, hsep = 0.6cm]{\color{black}\textbf{(a)}}{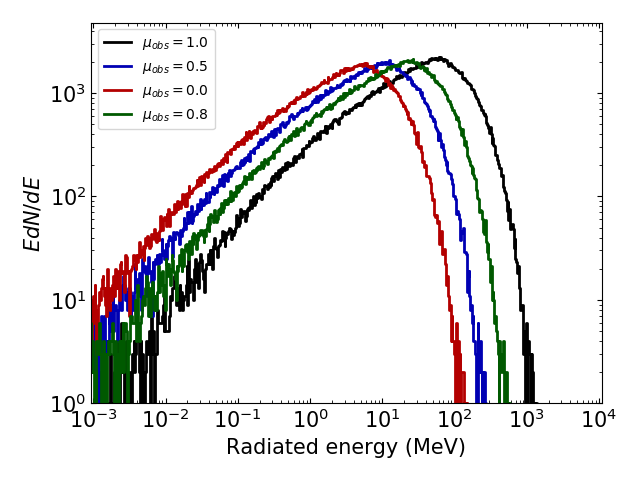} \\
    \subfigimg[width=\linewidth, pos = ur, vsep = 1 cm]{\color{black}\textbf{(b)}}{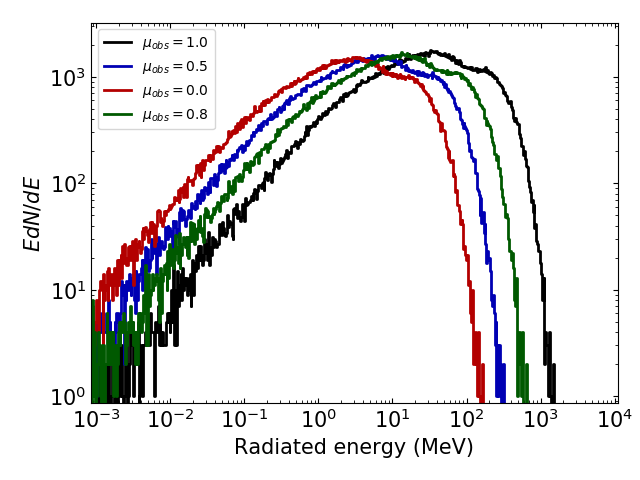} \\
     \subfigimg[width=\linewidth, pos = ur, vsep = 1 cm]{\color{black}\textbf{(c)}}{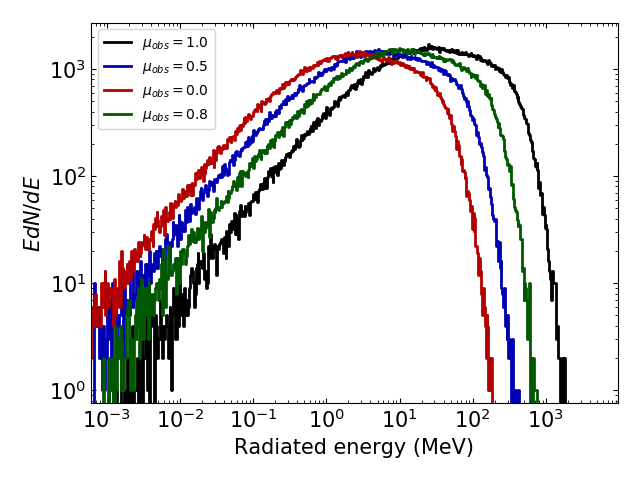}
  \end{tabular}
  \caption{Time-integrated angle-dependent Compton spectra in the sheath frame, resulting from simulations conducted at radiation temperature $\mathrm{\theta=10^{-5}}$ and different viewing angles: spectra are obtained at simulation time of (a) $\mathrm{t = 2000 ~\omega_{p,e}^{-1}}$, (b) $\mathrm{t = 4000 ~\omega_{p,e}^{-1}}$, and (c) $\mathrm{t = 5000 ~\omega_{p,e}^{-1}}$.}
  \label{fig:rad_spec_AD_UV}
\end{figure}

We observe sharp spectral peaks in the emissivity of Compton-scattered blackbody photons by mono-energetic relativistic electrons as evident in Figure \ref{fig:emisivty_eps_s}. Leveraging the pronounced spectral peaks, we apply a monochromatic approximation to convert the particle energy loss rate to a monochromatic radiation spectrum. This simplification significantly streamlines our calculations and reduces computational time, enabling more efficient analysis of the radiation spectrum. The obtained results, encompassing both angle-averaged and angle-dependent photon distributions, are presented in
\begin{figure}[!htb]
  \centering
  \begin{tabular}{c}
    \subfigimg[width=\linewidth, pos = ur, vsep = 1 cm]{\color{black}\textbf{(a)}}{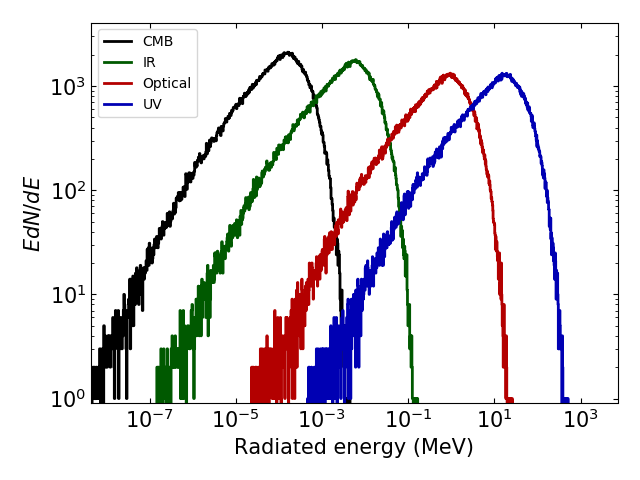} \\
    \subfigimg[width=\linewidth, pos = ur, vsep = 1 cm]{\color{black}\textbf{(b)}}{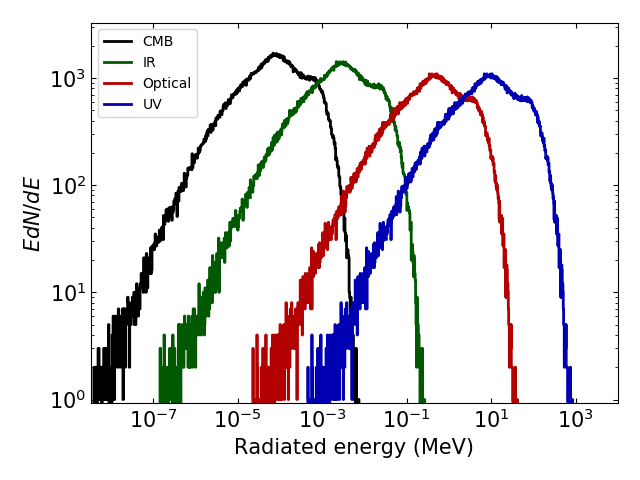} \\
     \subfigimg[width=\linewidth, pos = ur, vsep = 1 cm]{\color{black}\textbf{(c)}}{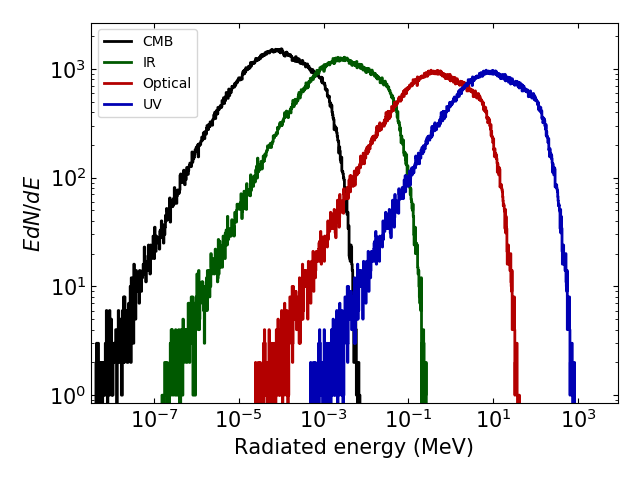} \\
  \end{tabular}
  \caption{Time-integrated angle-averaged Compton spectra in the sheath frame obtained from simulations conducted for different radiation temperatures of angle-averaged blackbody photon fields: the spectra correspond to three distinct phases: (a) early phase of the simulations at $\mathrm{t = 1500 ~\omega_{p,e}^{-1}}$, (b) intermediate phase at $\mathrm{t = 2500 ~\omega_{p,e}^{-1}}$, and (c) later steady state at $\mathrm{t = 3000 ~\omega_{p,e}^{-1}}$.}
  \label{fig:rad_spec_AI}
\end{figure}
\FloatBarrier
\noindent Figures \ref{fig:rad_spec_AI} and \ref{fig:rad_spec_AD_UV}.
\FloatBarrier
Throughout the various stages of the simulations, we observe distinct evolutions in the Compton spectra of relativistic electrons interacting with photon fields. In the early phases of the simulations, both angle-integrated and angle-dependent cases exhibit a quasi-thermal inverse Compton spectrum with a single dominant component. We observe a quasi-thermal inverse Compton spectrum with a single component around simulation time of $\mathrm{t = 1500/\omega_{p,e}}$ for an angle-integrated photon field and $\mathrm{t = 2000/\omega_{p,e}}$ for an angle-dependent photon distribution, as depicted in Figures \ref{fig:rad_spec_AI}(a) and \ref{fig:rad_spec_AD_UV}(a). As the simulation progresses, the Compton spectra undergo noteworthy changes. The single-component spectrum develops into a double-component spectrum, around $\mathrm{t = 2500/\omega_{p,e}}$ and $\mathrm{t = 4000/\omega_{p,e}}$ for the angle-integrated and angle-dependent photon distributions, respectively (see Figures \ref{fig:rad_spec_AI}(b) and \ref{fig:rad_spec_AD_UV}(b)). In the later steady state of the simulations, observed beyond $\mathrm{t = 3000/\omega_{p,e}}$ and $\mathrm{t = 5000/\omega_{p,e}}$, respectively, a distinct spectral pattern emerges. The Compton spectrum exhibits a quasi-thermal low-frequency spectrum with a cut-off power-law tail (see Figures \ref{fig:rad_spec_AI}(c) and \ref{fig:rad_spec_AD_UV}(c)).

The steady state is achieved later in the case of the inverse Compton spectra from scattering an angle-dependent photon field compared to the angle-averaged photon field. The angle-dependent Compton spectra exhibit a flatter peak in comparison to the angle-averaged Compton spectra. The observed peak energy in the angle-dependent radiation spectra is approximately $\mathrm{2.7 \theta \gamma^2}$ for $\mathrm{\cos (\theta_{obs}) = 1}$. As the value of $\mathrm{\cos(\theta_{obs}})$ decreases, the peak energy of the radiation spectra progressively shifts towards lower energies. This observed correlation between the peak energy and $\mathrm{\cos(\theta_{obs})}$ highlights the pronounced angular dependence intrinsic to the radiative output in relativistic jets. The radiation Spectra obtained in the case of an angle-dependent photon distribution exhibit similarities to those obtained for angle-averaged photons.

In our simulations, we observe the angle-averaged radiation spectra attaining a steady state beyond $\rm{t = 3000/\omega_{p,e}}$, while angle-dependent radiation spectra beyond $\rm{t = 5000/\omega_{p,e}}$. Beyond this stage, there are no significant fluctuations in their shapes, reaffirming the reliability and accuracy of our results and reflecting the dynamic equilibrium achieved in our simulated system. The time delay in reaching a steady state for the angle-dependent photon field's inverse Compton spectra, in comparison to the angle-averaged photon field, can be attributed to the increased complexity and larger parameter space resulting from the inclusion of angle-dependent photons. This introduces a higher computational demand for thorough sampling and evaluation of radiation outputs, leading to extended times needed to establish a stable equilibrium and achieve consistent radiation outputs. The broader and flatter spectral peak observed in the angle-dependent Compton spectra
\begin{figure}[!htb]
  \centering
  \begin{tabular}{c} 
    \subfigimg[width=\linewidth,pos=ul, hsep=2cm]{\textbf{(a)}}{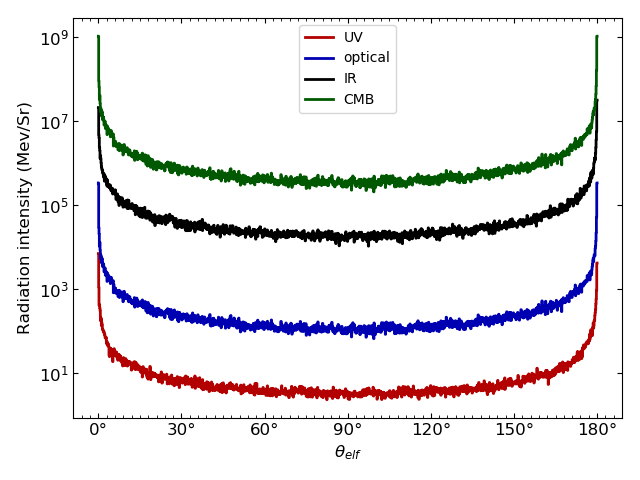} \\
    \subfigimg[width=\linewidth,pos=ul, hsep=2cm]{\textbf{(b)}}{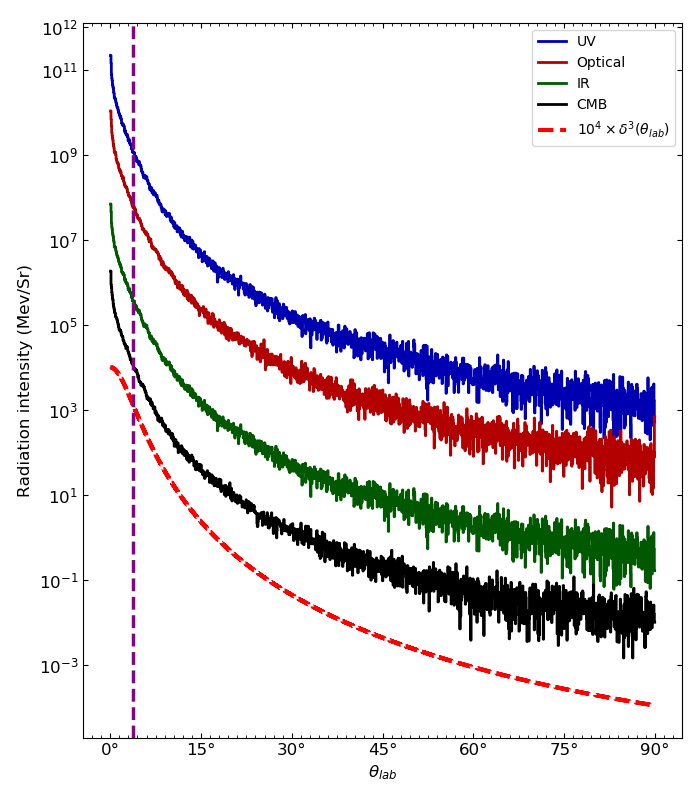} 
  \end{tabular}
  \caption{Radiation intensity resulting from Compton scattering of an angle-averaged photon field with varying temperatures as a function of the viewing angle of the jet at a simulation time of $\mathrm{t = 3000 \omega_{p,e}^{-1}}$: panel (a) displays the global radiative energy distribution per unit solid angle in the ELF, while panel (b) shows the same distribution in the laboratory frame (sheath). The dashed curve represents the $\delta^3$ boosting pattern, characteristic of co-moving isotropic photon emission. The violet dashed line represents $\mathrm{1/\Gamma}$.}
    \label{fig:rad_intensity_AI} 
\end{figure}
\FloatBarrier
\begin{figure}[!htb]
  \centering
  \begin{tabular}{c}
    \subfigimg[width=\linewidth,pos=ul, hsep = 2cm]{\textbf{(a)}}{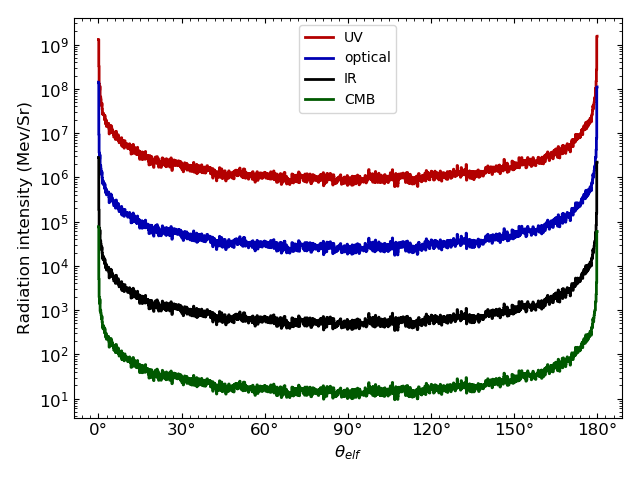} \\
    \subfigimg[width=\linewidth,pos=ul, hsep = 2cm]{\textbf{(b)}}{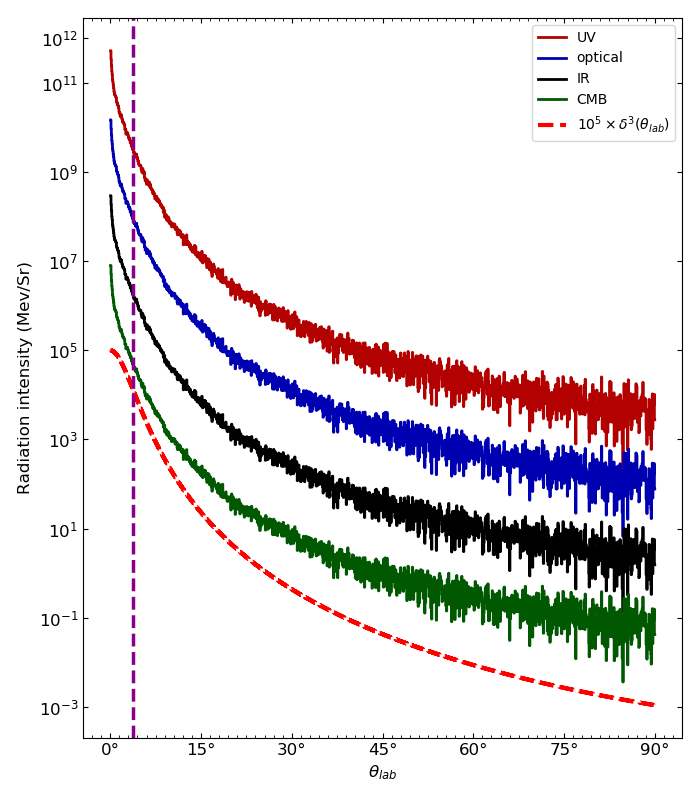} 
  \end{tabular}
  \caption{Radiation intensity of an angle-dependent photon field, subject to Compton upscattering by relativistic electrons as a function of the viewing angle of the jet at a simulation time of $\mathrm{t = 5000 \omega_{p,e}^{-1}}$ for varying radiation temperatures: panel (a) exhibits the global radiative energy distribution per unit solid angle in the ELF. Panel (b) shows the corresponding distribution in the laboratory frame. The dashed curve represents the $\mathrm{\delta^3}$ boosting pattern.}
    \label{fig:rad_intensity_AD}
\end{figure} 
\FloatBarrier
\noindent signifies a greater number of scattering events, indicating a more intricate interplay between photons and particles in this scenario. This difference in spectral characteristics further contributes to the increased time required to attain a steady state in the angle-dependent case.
\subsection{Observable Features of Radiation Spectra \label{subsec:obs_rad}}
To establish a connection between the radiation output obtained from the simulations and the observational data, we can plot the radiation intensity ($\mathrm{dE/d\Omega}$) as a function of the viewing angle. Here, $\mathrm{dE}$ represents the radiation energy and $\mathrm{d\Omega}$ denotes the differential solid angle, defined as $\mathrm{d\Omega = \sin\theta_{ph} d\theta_{ph} d\phi_{ph}}$. The angle $\mathrm{\theta_{ph}}$ represents the angle between the final direction of radiation and the axis of the jet (i.e., the viewing angle), while $\mathrm{\phi_{ph}}$ represents the azimuthal angle. The system is azimuthally symmetric in $\mathrm{\phi}$, so we may set $\mathrm{d\phi_{ph} = 2\pi}$. In the ELF, as depicted in Figures \ref{fig:rad_intensity_AI}(a) and \ref{fig:rad_intensity_AD}(a), the spine electrons emit the highest radiation intensities along the jet axis (i.e. at $\mathrm{\theta_{ph} = \theta_{elf} = 0^\circ}$).

In order to determine the radiation intensity as observed by an observer, we apply a Lorentz transformation into the sheath frame. Figures \ref{fig:rad_intensity_AI} and \ref{fig:rad_intensity_AD} show plots of the radiation intensity versus viewing angle and demonstrate that the inverse Compton radiation emitted from the jet's SBL experiences a strong boost in the forward direction, with a characteristic angle much smaller than $\mathrm{1/\Gamma}$. This is indicated by the violet dashed lines in Figures \ref{fig:rad_intensity_AI}(b) and \ref{fig:rad_intensity_AD}(b).

In the case of isotropic photon emission in the co-moving frame of an emission region moving with Lorentz factor $\mathrm{\Gamma}$ along the jet viewed at an angle $\mathrm{\theta_{lab}}$, the radiative energy flux $\mathrm{dF/dE}$ is boosted by a factor $\mathrm{\delta^3}$, with $\mathrm{\delta = 1/\Gamma(1 - \beta_\Gamma \cos \theta_{lab})}$ being the Doppler factor. The dashed curve depicted in Figures \ref{fig:rad_intensity_AI}(b) and \ref{fig:rad_intensity_AD}(b) illustrate the comparison of the simulated beaming characteristic to this $\delta^3$ pattern.
\section{Conclusion \label{sec:conclusion}}
In this article, we have presented the results of our study on the self-generated electric and magnetic fields and particle acceleration occurring at relativistic SBLs in relativistic jets associated with AGN and GRBs, and the subsequent inverse-Compton radiation output. Through the utilization of fully kinetic PiC simulations incorporating electron-ion plasma, we have shown that the electric and magnetic fields are self-generated and play a vital role in particle acceleration in SBLs. High energy particles are accelerated along the magnetic field lines, resulting in an anisotropic momentum distribution. This anisotropic particle distribution contributes effectively to the radiative output, along with the exchange of particles between the spine and sheath regions.

We employed the monochromatic approximation to compute the radiation spectra motivated by the sharp spectral peaks in the emissivity resulting from the Compton scattering of blackbody photons by monoenergetic relativistic electrons, as illustrated in Figure \ref{fig:emisivty_eps_s}. The difference between the monochromatic approximation and a full integration over the emission profiles is expected to be very small, as the spectral shape is dominated by the shape of the broad, thermal+non-thermal electron spectra and not the emissivity profiles of mono-energetic electrons. The recent work by \citep{2020Del_Gaudio} has introduced a promising numerical prospect to investigate radiative processes like inverse Compton emission. Their benchmarked algorithm seamlessly integrates with the standard PiC loop, leading to improved computational efficiency. We plan to experiment with this scheme in future work. The radiative output in our study has been directly obtained from the PiC simulation, from which we have calculated the radiation intensity to study the angle-dependent inverse-Compton spectra, self-consistently taking into account inverse-Compton cooling of relativistic electrons. To our knowledge, investigations into the angular dependence of radiative emissions from shear boundary layers (SBLs) within relativistic jets have not been conducted previously. Our research marks the pioneering effort to explore particle acceleration, the ensuing inverse Compton spectra, and the impact of inverse Compton cooling on the dynamics of relativistic electrons, while considering the impact of angular variations within these shear layers. We looked at both angle-averaged and angle-dependent distributions of target photons, using a $\mathrm{\delta}$-function approximation for the target photon field. Our findings showed that IC cooling can have a non-negligible impact on the acceleration of relativistic particles at SBLs, causing high-energy electrons to cool down to slightly lower energies. As expected, the inverse-Compton cooling effect becomes more pronounced for higher target-photon blackbody temperatures. In the early stages of simulations, both angle-integrated and angle-dependent inverse Compton radiation spectra exhibit a single component, quasi-thermal radiation spectrum. As the simulations progress, the spectrum evolves into a two-component spectrum, which eventually becomes a quasi-thermal low-frequency spectrum with a cut-off power-law tail.

The radiation emitted is strongly boosted along the jet axis, with a characteristic opening angle much less than 1/$\mathrm{\Gamma}$. This boosting is more powerful than what would be expected from Doppler boosting of an isotropic radiation field in the co-moving frame of the spine. The Doppler factor estimates for TeV blazars using one-zone models are inconsistent with observations made with VLBI. Our findings suggest that the extreme beaming patterns of particles accelerated at SBLs could resolve the long-standing problem of the Doppler factor crisis \citep[e.g.,][]{Lyutikov_2010}. In future work, we will explore the effects of a non-zero initial magnetic field on the plasma in relativistic jets, which could involve investigating a relativistic magnetically dominated electron-positron jet interacting with a weakly magnetized electron-ion ambient plasma. Recent work by \citet{Sironi_2021} has shown that such systems can develop Kelvin-Helmholtz instabilities and kinetic-scale reconnection layers, which provide a first-principles mechanism for particle injection into shear-driven acceleration. This could offer valuable insights into the complex dynamics of these shear-structured jets.

\acknowledgments{We would like to thank Anatoly Spitkovsky for initial discussions on the implementation of shear-driven particle acceleration in PiC codes. This work was supported through the South African Research Chair Initiative (SARChI) of the Department of Science and Technology and the National Research Foundation (NRF) of South Africa under SARChI chair grant no. 64789.}


\let\clearpage\relax
\def\bibsection{\section*{References}}
\bibliography{main.bib}
\bibliographystyle{aasjournal}


\end{document}